\documentclass[%
superscriptaddress,
showpacs,
 amsmath,amssymb,
 aps,prl,
floatfix,
]{revtex4-1}

\usepackage{graphicx}
\usepackage{dcolumn}
\usepackage{bm}
\usepackage{braket}
\usepackage{color}
\usepackage{subcaption}
\usepackage[font=footnotesize,labelfont=bf]{caption}
\usepackage[font=footnotesize,labelfont=bf]{subcaption}
\usepackage{hyperref}

\begin{document}

\preprint{APS/123-QED}

\title{Cavity QED engineering of spin dynamics and squeezing in a spinor gas}

\author{Stuart J. Masson}\email{smas176@aucklanduni.ac.nz} 
\affiliation{Dodd-Walls Centre for Photonic and Quantum Technologies, Department of Physics, University of Auckland, Private Bag 92019, Auckland 1142, New Zealand}
\author{M.~D. Barrett}
\affiliation{Centre for Quantum Technologies, 3 Science Drive 2, Singapore, 117543}
\affiliation{Department of Physics, National University of Singapore, 3 Science Drive 2, Singapore, 117543}
\author{Scott Parkins}\email{s.parkins@auckland.ac.nz}
\affiliation{Dodd-Walls Centre for Photonic and Quantum Technologies, Department of Physics, University of Auckland, Private Bag 92019, Auckland 1142, New Zealand}

\date{\today}

\begin{abstract}
We propose a method for engineering spin dynamics in ensembles of integer-spin atoms confined within a high-finesse optical cavity. Our proposal uses cavity-assisted Raman transitions to engineer a Dicke model for integer-spin atoms, which, in a dispersive limit, reduces to effective atom-atom interactions within the ensemble. This scheme offers a promising and flexible new avenue for the exploration of a wide range of spinor many-body physics. As an example of this, we present results showing that this method can be used to generate spin-nematic squeezing in an ensemble of spin-1 atoms. With realistic parameters the scheme should enable substantial squeezing on time scales much shorter than current experiments with spin-1 Bose-Einstein condensates.
\end{abstract}

\maketitle

Gases of ultracold Bose atoms possessing internal spin degrees of freedom -- spinor Bose gases -- offer a remarkable variety of possibilities for the investigation of quantum fluids, in contexts that include, for example, magnetism, superfluidity, and many-body quantum dynamics \cite{Kawaguchi12,StamperKurn13}. In this latter context, tremendous experimental progress has occurred in recent years based upon collision-induced spin-mixing dynamics in spinor Bose-Einstein condensates (BECs) \cite{Schmaljohann04,Chang04,Chang05,Kronjager05,Sadler06,Black07,Liu09,Leslie09,Klempt09,Klempt10,Bookjans11PRL1,Lucke11,Gross11,Hamley12,Hoang13,Peise15,Hoang16NatComm,Anquez16,Hoang16PNAS,Linnemann16,Kruse16,Luo17}. Such systems have allowed for the generation of quantum spin squeezing and entangled states \cite{Lucke11,Gross11,Hamley12,Hoang13,Peise15,Hoang16NatComm,Linnemann16,Kruse16,Payrits16,Huang17,Luo17} following a range of proposals \cite{Pu00,Duan02,Mustecaplioglu02,Sau10,Zhang13,Sun17}, as well as the study of quantum phase transitions \cite{Sadler06,Black07,Liu09,Anquez16,Hoang16PNAS,Luo17} and the parametric amplification of quantum spin fluctuations \cite{Leslie09,Klempt09,Klempt10,Scherer10,Hoang16NatComm}.

Spinor BECs in which atoms in all magnetic sub-levels of a single hyperfine ground state (e.g., the $F=1$ ground state of ${}^{87}\textrm{Rb}$) are condensed correspond to ensembles of integer-spin particles. 
For small, tightly confined condensates, one may assume that the different atomic states have the same spatial wave function -- the single-mode approximation -- after which one can show that the collisional spin dynamics is described by a Hamiltonian of the form $\lambda \hat{\bf S}^2$, where $\hat{\bf S}=(\hat S_x,\hat S_y,\hat S_z)$ is the total spin vector (operator) and $\lambda$ is the collisional spin interaction energy per particle integrated over the condensate \cite{Law98PRL,Pu99}. The spinor dynamical rate is $c=2N\lambda$, where $N$ is the number of atoms, and is typically on the order of 10 Hz for $40,000$ $^{87}$Rb atoms \cite{Hamley12}. If the longitudinal magnetization $\braket{ \hat S_z}$ is a constant of the motion (e.g., zero), then this Hamiltonian can be reduced further to $\lambda (\hat S_x^2+\hat S_y^2)$. With the addition of a magnetic field, the Hamiltonian gains a linear Zeeman shift $p\hat{S}_z$ (which can also be assumed to be a constant of the motion) and a quadratic Zeeman shift $q \hat{N}_0$, where $\hat{N}_0$ is the population in the $m=0$ state. The ratio $q/c$ describes a rich phase diagram, with highly entangled ground states in several limits \cite{Zhang13,Hoang16PNAS,Luo17,Sun17}. In particular, if $|q| \ll c > 0$, the ground state is the spin singlet state, which has fundamental interest due to its high entanglement, but also applications ranging from precision measurements \cite{UrizarLanz13} to no-classical solution quantum information processing \cite{Cabello02}. In addition, the transitions between these different phases are of interest with respect to the Kibble-Zurek mechanism \cite{Damski07,Saito07,Anquez16}.

In this Letter, we propose an alternative scheme to producing spin-mixing dynamics in a gas of integer-spin atoms that uses cavity-mediated Raman transitions to engineer the required spinor dynamics. Our proposal borrows from earlier schemes for engineering effective Dicke models of collective two-level-atom ensembles coupled strongly to a quantized cavity mode \cite{Dimer07,Morrison08PRL,Morrison08PRA}, but considers an arguably simpler configuration and limit, which yields a Dicke model for integer-spin (alkali) atoms. This approach has in fact been demonstrated very recently in a study of non-equilibrium phase transitions in this model \cite{Zhiqiang17}. In the dispersive limit of this model, where the cavity mode is only virtually excited, the resulting Hamiltonian mimics collisional interactions in a spinor BEC.

\begin{figure}[b]
\includegraphics[width=0.5\textwidth]{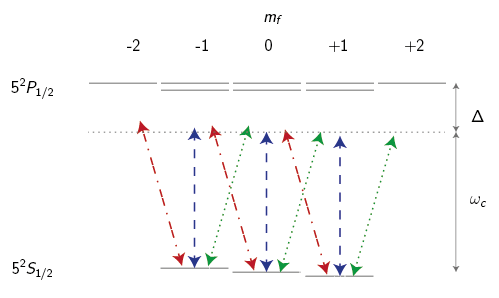}
\caption{Level diagram for the implementation of an effective Dicke model using the $F=1$ ground state of ${}^{87}{\rm Rb}$. Interactions are engineered via Raman transitions on the $D_1$ line mediated by a cavity mode (blue dashed line), and $\sigma_+$ (green dotted) or $\sigma_-$ (red dot-dashed) polarised laser fields.\label{diagram}}
\end{figure}

We consider an ensemble of alkali atoms tightly confined (e.g. by a three-dimensional optical lattice) inside a high-finesse optical cavity. The atomic ensemble is considered dilute enough to exclude direct atom-atom interactions, while the atoms are coupled uniformly to cavity and laser fields. As illustrated in Fig.~\ref{diagram}, we consider a scheme of cavity-assisted Raman transitions in which the fields are very far detuned from the relevant excited state manifold. Here, instead of isolating effective spin-1/2 systems \cite{Grimsmo13PRA}, we consider transitions within a complete hyperfine ground state; in this instance, the $F=1$ ground state of ${}^{87}{\rm Rb}$. Adiabatic elimination of the atomic excited states then creates an effective model for an ensemble of spin-1 atoms coupled to a cavity mode (see Supplementary Material).

In the limit that the detunings of the fields are very large -- in particular, much larger than the energy separations of the excited state hyperfine levels (e.g., as in \cite{Zhiqiang17}, where the detuning is 127~GHz) -- then the internal structure of the excited state manifold becomes unimportant, and symmetries in the dipole operator cause the effective Hamiltonian to simplify greatly. Considering an open quantum system, with the cavity field decay rate given by $\kappa$ (but atomic spontaneous emission neglected due to the large detuning), this model is described by the master equation for the atom-field density operator $\rho$,
\begin{equation}
\dot{\rho} = -i[\hat{H},\rho] + \kappa\mathcal{D}[\hat{a}]\rho ,
\end{equation}
where $\hat{a}$ is the cavity mode annihilation operator, $\mathcal{D}[\hat{a}]\rho = 2\hat{a}\rho \hat{a}^\dagger - \rho \hat{a}^\dagger \hat{a} - \hat{a}^\dagger \hat{a} \rho$, and 
\begin{align}
\hat{H} &= \omega \hat{a}^\dagger \hat{a} + \omega_0 \hat{S}_z \nonumber
\\
& + \frac{\lambda_-}{\sqrt{2N}} (\hat{a}\hat{S}_+ + \hat{a}^\dagger \hat{S}_-) + \frac{\lambda_+}{\sqrt{2N}} (\hat{a}\hat{S}_- + \hat{a}^\dagger \hat{S}_+).
\end{align}
Here, $\hat{S}_\pm$ are the spin-1 collective raising and lowering operators, while the coefficients of the various terms (for the $F=1$ manifold  in $^{87}$Rb coupled via the $D_1$ line) are given by
\begin{align}
\omega &= \omega_c - \frac{\omega_- + \omega_+}{2} + \frac{Ng^2}{3\Delta} , \\
\omega_0 &= \omega_z - \frac{\omega_- - \omega_+}{2} + \frac{\Omega_-^2-\Omega_+^2}{24\Delta} , \\
\lambda_\pm &= \frac{\sqrt{N} g\Omega_\pm}{12\Delta}.
\end{align}
Here $\omega_{c}$ is the frequency of the cavity mode, $\omega_\pm$ ($\Omega_\pm$) are the bare frequencies (Rabi frequencies) of the $\sigma_\pm$ polarised laser fields, $\omega_z$ is the Zeeman splitting of the $F=1$ levels (due to an applied magnetic field, if present), $g$ is the single-atom-cavity coupling strength (for the ${}^{87}{\rm Rb}$ $D_2$ line cycling transition), and $\Delta$ is the detuning of the fields from the atomic resonance.

This configuration provides a ``clean'' and tuneable system with which to study the Dicke model, as demonstrated in \cite{Zhiqiang17}. In particular, it has the independence of couplings $\lambda_\pm$ not present in BEC formulations of the Dicke model \cite{Baumann10, Baumann11}, while it also avoids a non-linear coupling term of the form $\hat{S}_z \hat{a}^\dagger \hat{a}$ that features in all the current spin-1/2 versions of the Dicke model \cite{Dimer07,Nagy10,Keeling10,Bhaseen12,Baumann10,Baumann11,Grimsmo13PRA}. We note that it can also be applied to other hyperfine ground states in alkali atoms, enabling, e.g., tuneable interactions for ensembles of effective spin-2 (${}^{87}{\rm Rb}$ or ${}^{85}{\rm Rb}$), 3 (${}^{85}{\rm Rb}$, ${}^{133}{\rm Cs}$), or 4 (${}^{133}{\rm Cs}$) atoms.

While most previous work has considered many-body cavity QED with two-level systems \cite{Ritsch13,Noh17}, the generalization to integer spin ensembles offers a range of interesting physics not available to spin-1/2 systems. Integer spins have more degrees of freedom, which means that there are different ways to manipulate excitations and constrain the state. In particular, a coherent ensemble of integer spins is not limited to the surface of the angular momentum Bloch sphere. This allows for novel entangled states such as the spin-singlet, two-mode squeezed spin states or, as discussed in more detail below, the redistribution of quantum noise into degrees of freedom that are simply not present in two-level systems \cite{Vitagliano11,Vitagliano14,Behbood14}.

We now consider the dispersive limit in which the Raman transitions are themselves off-resonant, i.e., $\omega \gg \omega_0, \lambda_\pm$, in which case we can also adiabatically eliminate the cavity mode to yield the reduced master equation
\begin{equation}
\dot{\rho} = -i[\hat{H},\rho] + \frac{\kappa}{2N(\omega^2+\kappa^2)} \mathcal{D}[\lambda_- \hat{S}_- + \lambda_+ \hat{S}_+]\rho \label{masterequation},
\end{equation}
with
\begin{align}
\hat{H} &= \left[ \omega_0 - \frac{\omega(\lambda_-^2 - \lambda_+^2)}{2N(\omega^2+\kappa^2)} \right] \hat{S}_z \nonumber
\\
- &\frac{\omega}{2N(\omega^2+\kappa^2)} \left[ (\lambda_- + \lambda_+)^2 \hat{S}_x^2 + (\lambda_- - \lambda_+)^2 \hat{S}_y^2  \right].
\end{align}
If we set $\lambda_+ = 0$ and $\lambda_- = \lambda$ then (\ref{masterequation}) becomes
\begin{equation}\label{eq:ME}
\dot{\rho} = -i[\hat{H},\rho] + \frac{\Gamma}{2N} \mathcal{D}[\hat{S}_-] \rho,
\end{equation}
where
\begin{equation}
\hat{H} = \omega_0^\prime \hat{S}_z + \frac{\Lambda}{2N} (\hat{S}_x^2 + \hat{S}_y^2) ,
\end{equation}
with parameters given by
\begin{align}
\omega_0^\prime = \omega_0 + \frac{\Lambda}{2N} , ~~~
\Lambda = -\frac{\omega \lambda^2}{\omega^2+\kappa^2} , ~~~
\Gamma = -\frac{\kappa}{\omega} \Lambda .
\end{align}

Note that, by choosing the sign of $\omega$, it is possible to produce ferromagnetic or anti-ferromagnetic behaviour with the same atomic species. An artificial quadratic Zeeman shift could also be added to the system by, for example, a weak $\pi$-polarised laser field acting near the $F' = 1$ line in the excited manifold. Then, in the limit that $\Gamma / \Lambda \ll 1$, the atoms will undergo spin-mixing interactions with dynamics of the sort found in spinor BECs. However, here the relevant dynamical rate is set by Raman transition rates, light shifts and detunings, and can therefore be orders of magnitude larger than in spinor BECs. Consider, e.g., the feasible experimental parameters $\{ g,\kappa ,\gamma\}/(2\pi ) = \{ 10,0.2,6\}$~MHz (see, e.g., \cite{Sames14,Reimann15,Zhiqiang17}), where $\gamma$ is the atomic spontaneous emission linewidth. With $N=10^4$ atoms, values of $\lambda /(2\pi )\simeq 200$~kHz are then readily achievable, which, with $\omega/(2\pi)\simeq 4$~MHz, lead to $\Lambda/(2\pi)\simeq 10$~kHz and $\Gamma = 0.05 \Lambda$. This means that such a system can emulate the dynamics of a spinor BEC, but orders of magnitude faster.  

This Hamiltonian gives the opportunity to study a range of models, such as the Lipkin-Meshkov-Glick model, where, unlike in the spin-1/2 case, the spin-1 case features quantum chaotic behaviour \cite{Grass13}. Similar studies have shown that spin-2 models can offer very different dynamics again \cite{Kronjager08}.

The methods described above are not limited to emulations of spinor BEC physics. Since this is an open system, there is also flexibility to deliberately engineer particular dissipative evolution or monitor the cavity output to gain information about the evolution without destroying the spinor gas.

It is also possible to produce Hamiltonians which do not naturally arise in spinor BECs. For example, by setting $\lambda_- = \lambda_+$ we obtain a Hamiltonian $\sim \hat{S}_x^2$, which produces squeezing via one-axis twisting in two-level systems \cite{Kitagawa93,Gross10,Riedel10,Ockeloen13,Muessel14,Leroux10}. By adding more cavity and laser modes an even wider range of Hamiltonians is possible \cite{Morrison08PRA,Yu16}, with, for example, the possibility of a two-axis twisting Hamiltonian $\propto \hat{S}_x^2 - \hat{S}_y^2$ \cite{Kitagawa93}, which can offer Heisenberg limited metrology, but has yet to be implemented experimentally. Such systems with spin-1 (or higher) particles offer the same squeezing possibilities, but should also allow further novel, many-body ground states and dynamical phenomena.

The principle behind our scheme could also be applied to the emerging field of quantum simulation with cold atoms coupled to a photonic crystal waveguide \cite{Douglas15}. Atoms coupled to the waveguide, but with the atomic resonance frequency located within a photonic band-gap, enable localized excitations at the atom trapping sites, while tunneling of excitations between neighboring sites produces effective atom-atom interactions. While work to date has focussed on spin-1/2 systems, application of our approach should enable generalization of this work to integer-spin lattice models with engineered interactions that could be tuned in form, strength and range, allowing, for example, exploration of Haldane physics \cite{Haldane83}.

Now, we consider an example of how our scheme can be used to emulate spinor BEC physics. In particular, we consider model (\ref{eq:ME}) and the preparation of ``spin-nematic squeezing'' in an ensemble of spin-1 atoms that are initially prepared in the $m=0$ sublevel \cite{Hamley12,HuangY15}. With a suitable choice of microscopic parameters, it is possible to set $\omega_0^\prime = 0$ (or at least approximately so). We note, however, that initially $\langle\hat{S}_z\rangle = 0$, and since $\hat{S}_z$ is conserved by the Hamiltonian, this term should not have any significant impact on the evolution, provided that the cavity-mediated damping of the spin (which does not conserve $\hat{S}_z$) is weak, i.e., $\kappa \ll \omega$, which means $\Gamma\ll\Lambda$. With this condition met, the Hamiltonian is an active generator of spin-nematic squeezing.

Spin squeezing is a well established method to produce metrological enhancement (for reviews, see \cite{Ma11,Pezze16}). In particular, atom interferometers can be used for precision measurements of acceleration, time, rotation, and, potentially, even gravitational waves \cite{Tino14}. If the input states to these interferometers are uncorrelated states of $N$ atoms, then the precision of the measurement is limited by the standard quantum limit, which scales as $1/\sqrt{N}$. However, by generating suitable entanglement within the atomic ensemble, it is possible to exceed this and approach the Heisenberg limit, which scales like $1/N$.

Considering ensembles of $N$ two-level, or spin-1/2, atoms with internal spin degrees of freedom, spin squeezing involves a redistribution of quantum noise on the angular momentum Bloch sphere in such a way as to produce reduced quantum fluctuations along one coordinate axis. Integer spin systems possess additional degrees of freedom associated with the quadrupole or nematic tensor operator $\hat Q_{ij} = \sum_{n=1}^N \hat{S}_i^{(n)} \hat{S}_j^{(n)} + \hat{S}_j^{(n)} \hat{S}_i^{(n)} - (4/3) \delta_{ij}$ where $(\{ i,j\}\in\{ x,y,z\})$, $\hat{S}_i^{(n)}$ are spin-1 angular momentum operators for a single atom, and $\delta_{ij}$ is the Kronecker delta function. Spin-nematic squeezing involves the redistribution of quantum noise in the subspaces $\{ \hat S_x,\hat Q_{yz},\hat Q_{zz} - \hat Q_{yy}\}$ and $\{ \hat S_y,\hat Q_{xz},\hat Q_{zz} - \hat Q_{xx}\}$ \cite{Hamley12}. Focussing on the first of these subspaces, the degree of spin-nematic squeezing can be characterised by a parameter $\xi_x$, which gives the metrological precision relative to the SQL for Ramsey interferometry and is calculated by minimising the following expression over the angle $\theta$,
\begin{equation}
\xi_x^2 = \frac{2\braket{[\Delta(\hat S_x\cos\theta + \hat Q_{yz}\sin\theta )]^2}}{|\braket{\hat Q_{zz}-\hat Q_{yy}}|} ,
\end{equation}
with $\xi_x^2<1$ indicating spin-nematic squeezing.

Fig.~\ref{squeezing} shows the development of spin-nematic squeezing with and without damping for an ensemble of $N=120$ atoms. These results are obtained from quantum trajectory simulations of the master equation (\ref{eq:ME}), in which we make use of a representation in terms of bosonic mode operators,  $\hat{a}_{m}$, for the three Zeeman states $m=0,\pm 1$ (see Supplementary Material). In particular, $\hat{a}_{m}\,(\hat{a}_{m}^\dagger)$ annihilates (creates) an atom in state $m$, and, e.g., $\hat{S}_-=\sqrt{2}(\hat{a}_0^\dagger\hat{a}_1+\hat{a}_{-1}^\dagger\hat{a}_0)$. Note that in this picture the Hamiltonian contains a term proportional to $\hat{a}^\dagger_0 \hat{a}^\dagger_0 \hat{a}_{-1} \hat{a}_{+1} + \hat{a}^\dagger_{-1} \hat{a}^\dagger_{+1} \hat{a}_0 \hat{a}_0$, which highlights explicitly the link to squeezing via four-wave mixing in light \cite{Hamley12,HuangY15}.

\begin{figure}[t]
\includegraphics[width=0.5\textwidth]{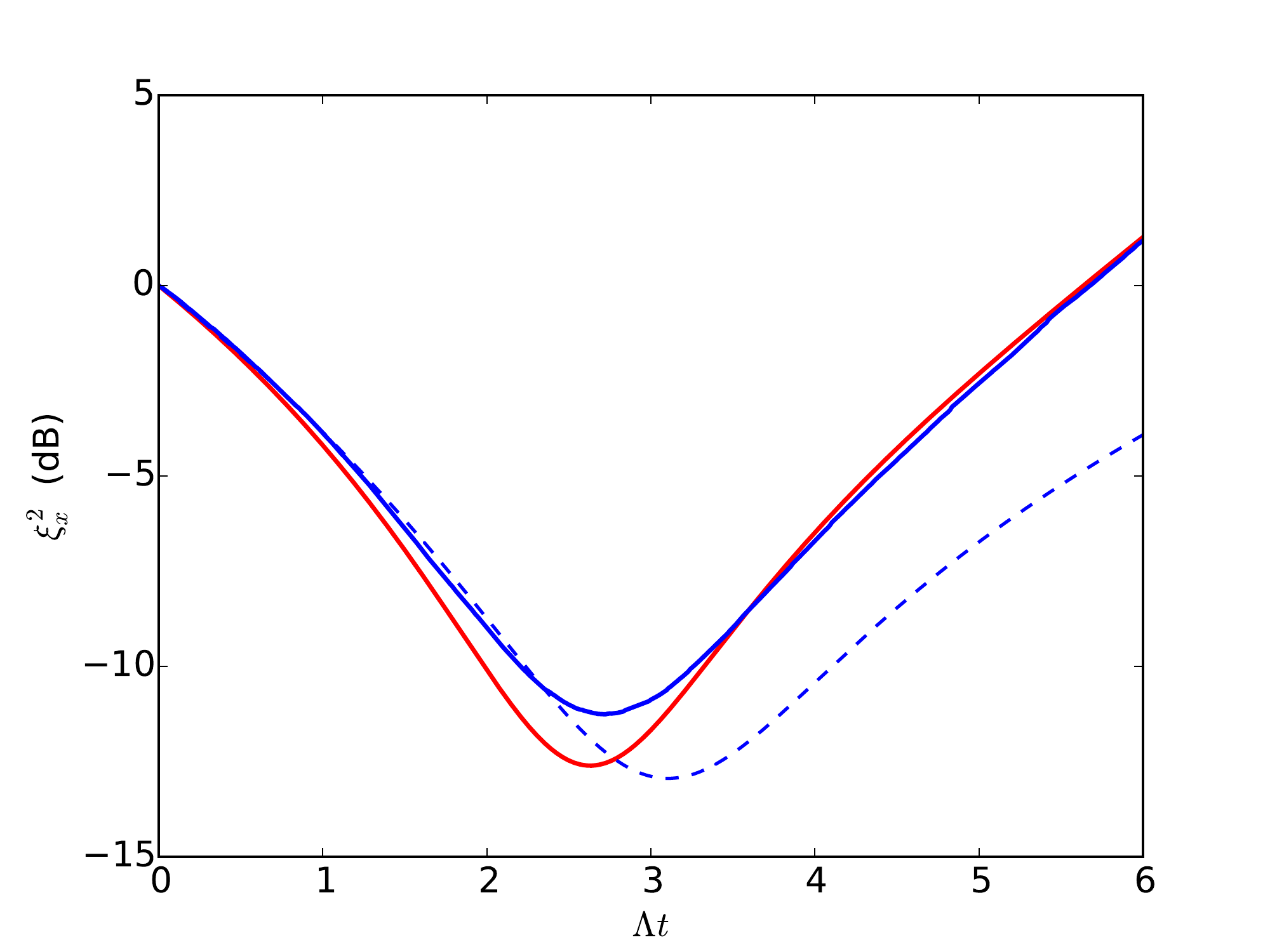}

\includegraphics[width=0.23\textwidth]{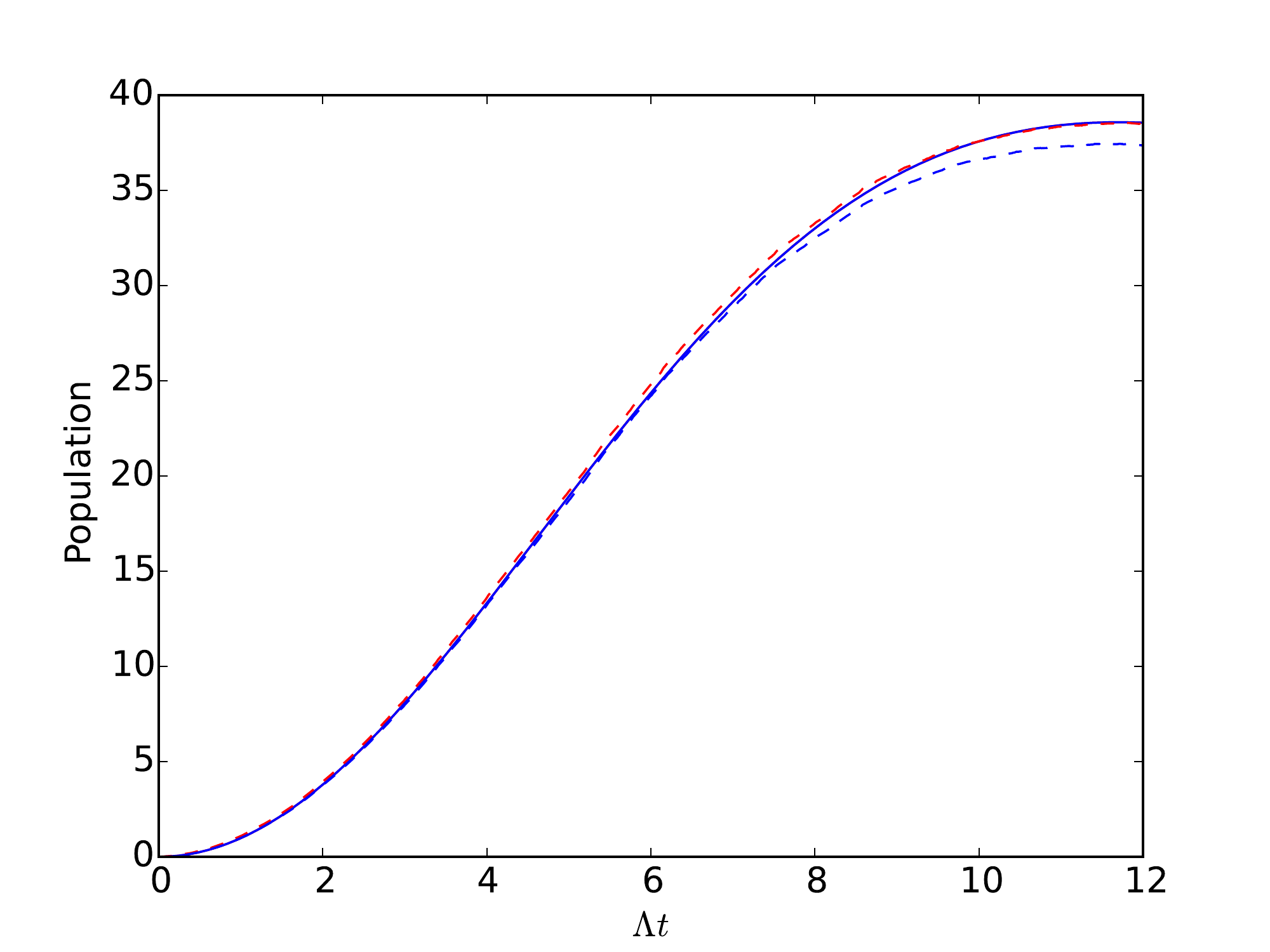}
\includegraphics[width=0.23\textwidth]{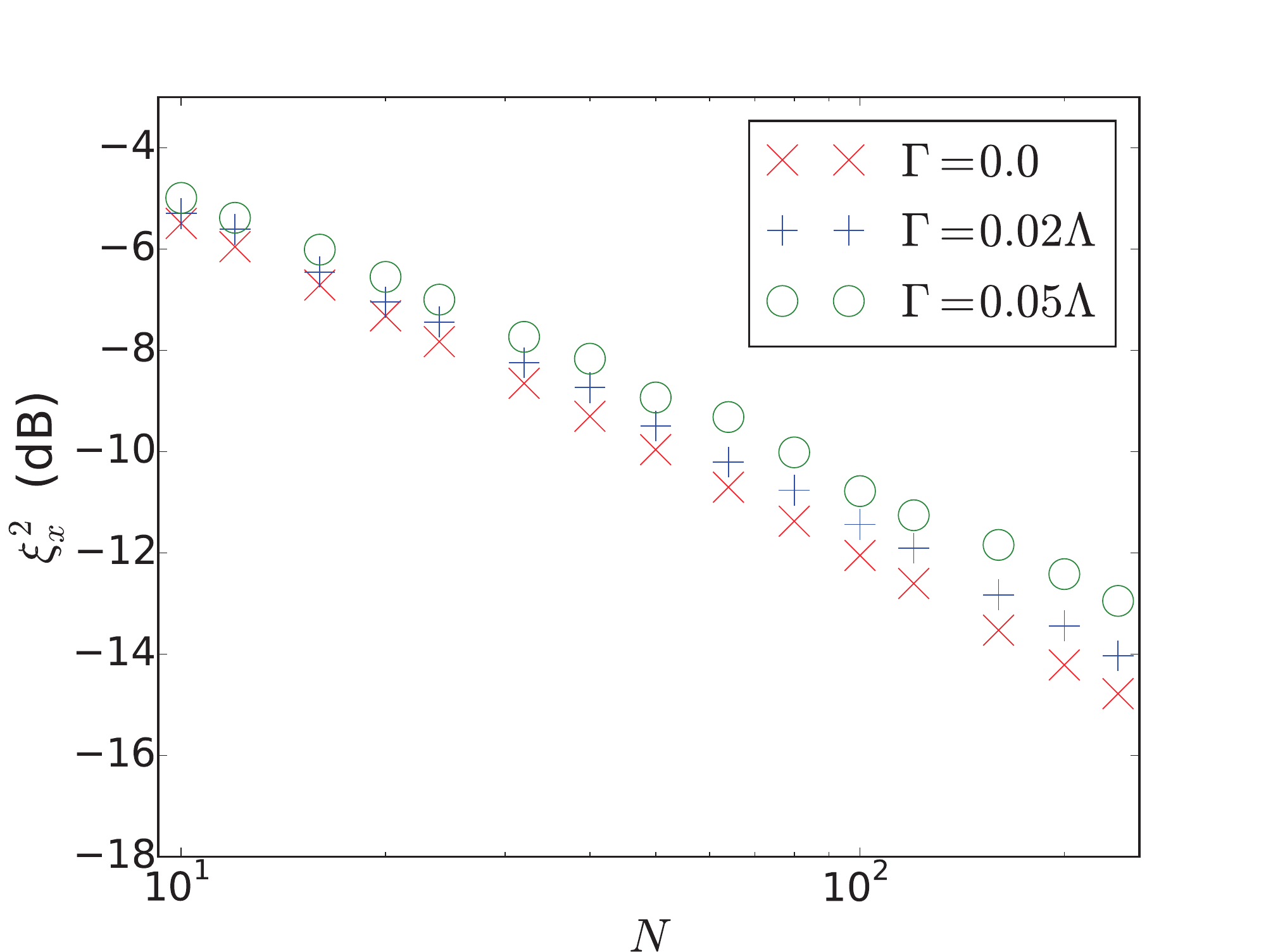}

\caption{(top) Time evolution of the spin-nematic squeezing for $N=120$ atoms without damping (red line) and, with damping rate $\Gamma = 0.05\Lambda$, an ensemble average of 1000 trajectories (dark blue) and a single trajectory in which no jumps occur (dashed dark blue). The phase angle in each case is just below $\theta=170^\circ$. (bottom left) Populations in each of the states $m=\pm 1$ for $\Gamma=0$ (solid line) and $\Gamma = 0.05\Lambda$ (dashed lines). (bottom right) Optimised squeezing scaling with atom number with and without damping. With damping, ensemble averages of 1000 trajectories were used to estimate the master equation result.\label{squeezing}}
\end{figure}

With $\Gamma = 0$ the system simply follows the Hamiltonian evolution and we see significant squeezing generated on a timescale $(\Lambda/2)^{-1}$. After this time, squeezing reduces and ultimately turns into anti-squeezing. This turnaround correlates with a growing number of atoms in the $m=\pm 1$ states, resulting in a reduction in $|\hat{Q}_{zz} - \hat{Q}_{yy}|$.

With the addition of a small rate of damping, which causes (infrequent) quantum jumps with $\hat{S}_-$, we find that the trajectories can be split into two categories. Firstly, those that reach the point of peak squeezing without a jump having occurred. Interestingly, these have squeezing at a slightly higher level than with $\Gamma=0$, meaning that the null measurement back action (which essentially adds an imaginary element to the spin-nematic squeezing generator) actually improves the squeezing. Secondly, when there is a jump before that point, the squeezing is substantially reduced, and so, on average, the presence of damping does decrease the degree of squeezing. However, for sufficiently small $\kappa/\omega$ such jumps should be rare. In addition, since these jumps are mediated by the cavity mode, i.e., they correspond to the emission of a photon from the cavity mode, then by monitoring that output and post-selecting based on the absence of a photon measurement, it would be possible to remove some of the runs with non-optimal squeezing (allowing for finite detection efficiency).

Fig.~\ref{squeezing} also illustrates more clearly how the best achievable squeezing varies with the number of atoms, with results obtained from trajectory simulations of (\ref{eq:ME}). For $\Gamma=0$ we find that $\xi_x^2\sim N^{-0.673}$ in the range of atom number that we consider. This indicates that this spin-nematic squeezing scales very similarly to one-axis twisting (where the squeezing scales at best as $N^{-2/3}$). Note that we have also considered spin-nematic squeezing in spin-2 particles, as would be relevant to the situation in which the present scheme is applied to the $F=2$ ground states in either ${}^{85}{\rm Rb}$ or ${}^{87}{\rm Rb}$, and find similar results.

\begin{figure}[t]
\includegraphics[width=0.5\textwidth]{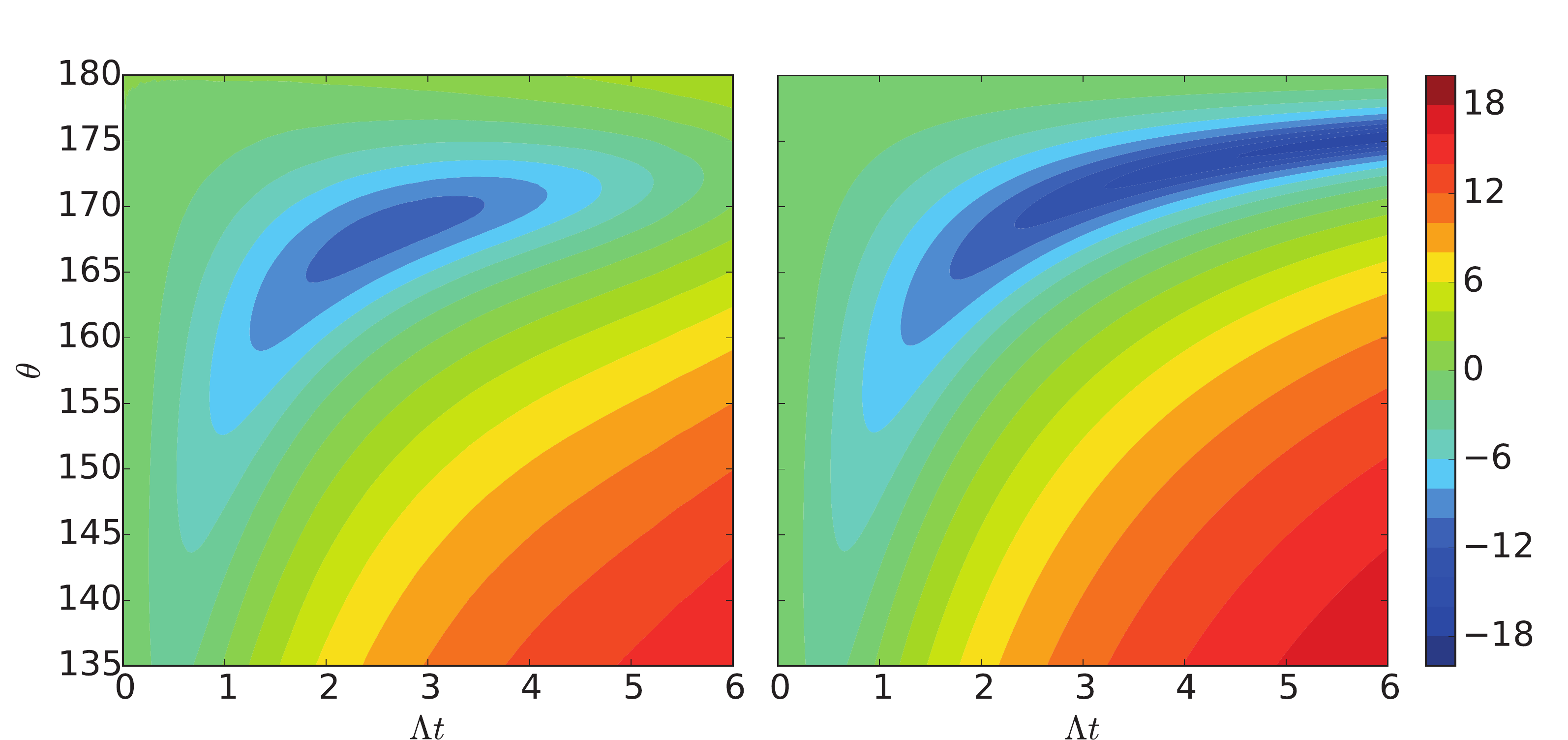}
\caption{Values of $\xi_x^2$ (in dB) as a function of time and phase angle for $\Gamma/\Lambda = 0.05$ with $N=120$ atoms (left) and in the undepleted mode approximation ($N\rightarrow\infty$), Eq.~(\ref{eq:xisq_Ninf}) (right).}
\label{xisqcontour}
\end{figure}

In Fig.~\ref{xisqcontour} we plot the squeezing parameter as a function of both time and phase angle $\theta$ for $N=120$ atoms (left) and in the limit of large $N$ (right), where we assume that the $m=0$ state is essentially undepleted and $\hat{a}_0$ can be replaced by $\sqrt{N}$. In this case it is possible to derive the simple result (see Supplementary Material)
\begin{align}\label{eq:xisq_Ninf}
\xi_x^2 = \left(\cos\theta + 2\Lambda t\,\sin\theta \right)^2 + (1+2\Gamma t)^2\sin^2\theta .
\end{align}
This agrees quite well with the $N=120$ results up to $\Lambda t\simeq 2$, but then predicts continued improvement in the degree of squeezing at longer evolution times and for phase angles that approach (slowly) $180^\circ$.

Finally, we note that the rate of atomic spontaneous emission due to off-resonant excitation of the $5{}^2P_{1/2}$ state is estimated, for our configuration, as $\Gamma_{\rm sp}=\gamma (\Omega^2/12\Delta^2)$, which gives $\Gamma_{\rm sp}/(\Lambda/2)\simeq 48\omega/(NC\kappa )$, where $C=2g^2/(\kappa\gamma)$ is the single atom cooperativity. For the parameters discussed above this ratio is $\sim 0.0006$. With more atoms and/or increased cooperativity, this can evidently be reduced even further.

To conclude, we have a proposed a method for engineering spinor dynamics using cavity-mediated Raman transitions and demonstrated that such a scheme could be used to produce spin-nematic squeezing in an ensemble of spin-1 atoms. We believe this work opens up a range of exciting possibilities for emulating spinor BEC dynamics on much shorter timescales and extending this to explore a much wider range of spinor physics with significant flexibility.

S.J.M. and S.P. thank Blair Blakie, Danny Baillie and Luke Symes for helpful conversations about spinor BECs and acknowledge support from the Marsden Fund of the Royal Society of New Zealand (Contract No. UOA1328). They also acknowledge the contribution of NeSI high-performance computing facilities to the results of this research. NZ's national facilities are provided by the NZ eScience Infrastructure and funded jointly by NeSI's collaborator institutions and through the Ministry of Business, Innovation and Employment's Research Infrastructure programme.

\newpage

\title{Supplementary information: Cavity QED engineering of spin dynamics and squeezing in a spinor gas}

\author{Stuart J. Masson}\email{smas176@aucklanduni.ac.nz} 
\affiliation{Dodd-Walls Centre for Photonic and Quantum Technologies, Department of Physics, University of Auckland, Private Bag 92019, Auckland 1142, New Zealand}
\author{M.~D. Barrett}
\affiliation{Centre for Quantum Technologies, 3 Science Drive 2, Singapore, 117543}
\affiliation{Department of Physics, National University of Singapore, 3 Science Drive 2, Singapore, 117543}
\author{Scott Parkins}\email{s.parkins@auckland.ac.nz}
\affiliation{Dodd-Walls Centre for Photonic and Quantum Technologies, Department of Physics, University of Auckland, Private Bag 92019, Auckland 1142, New Zealand}

\date{\today}

\maketitle

\section{Deriving the Dicke model for spin-1 particles in $^{87}$Rb}

Our proposed scheme is applicable to any magnetic manifold of an alkali metal, but here we show the full derivation for a specific example: the $F=1$ manifold of ${}^{87}$Rb. We consider an ensemble of cold atoms dilute enough to assume that direct atom-atom interactions are negligible. We take detuned Raman transitions to be driven on the $D_1$ line by counter-propagating $\sigma_+$ and $\sigma_-$ polarised lasers and a $\pi$ polarised cavity mode, and assume uniform-strength couplings of the atoms to all fields. This then gives us a model of indistinguishable atoms interacting identically with the light modes. We can thus derive the single atom model and then sum over the other atoms.

We first set up the full Hamiltonian. We use notation such that $\ket{m}$ represents a level $m$ in the $F=1$ ground manifold and $\ket{F',m_{F'}}$ represents a level $m_{F'}$ in the $F'$ excited hyperfine level. This Hamiltonian is then the sum of the following components ($\hbar = 1$),
\begin{align}
\hat{H}_{\mathrm{0}} &= \omega_c \hat{a}^\dagger \hat{a} + \omega_z (\ket{+1}\bra{+1} - \ket{-1}\bra{-1}) , \\
\hat{H}_{\mathrm{e}} &= \sum\limits_{F'} \sum\limits_{m_{F'}} \omega_{F'} \ket{F',m_{F'}}\bra{F',m_{F'}} , \\
\hat{H}_{\mathrm{c}} &= \frac{g}{\sqrt{12}} \hat{a} \left( \ket{1,-1}\bra{-1} + \sqrt{3} \ket{2,-1}\bra{-1} + 2 \ket{2,0} \bra{0} - \ket{1,+1}\bra{+1} + \sqrt{3}\ket{2,+1}\bra{+1}\right) + \mathrm{H.c.} , \\
\hat{H}_{\mathrm{+}} &= -\frac{\Omega_+\mathrm{e}^{-i\omega_+t}}{2\sqrt{12}} \left( \ket{1,0}\bra{-1} + \ket{2,0}\bra{-1} + \ket{1,+1}\bra{0} +  \sqrt{3}\ket{2,+1}\bra{0} + \sqrt{6} \ket{2,+2}\bra{+1}\right) + \mathrm{H.c.} , \\
\hat{H}_{\mathrm{-}} &= \frac{\Omega_-\mathrm{e}^{-i\omega_-t}}{2\sqrt{12}} \left( - \sqrt{6} \ket{2,-2}\bra{-1} + \ket{1,-1}\bra{0} - \sqrt{3} \ket{2,-1}\bra{0} + \ket{1,0}\bra{+1} - \ket{2,0} \bra{+1}\right) + \mathrm{H.c.}
\end{align}
Here $\hat{a} (\hat{a}^\dagger)$ is the annihilation (creation) operator for the cavity mode, $\{\omega_c,\omega_\pm,\omega_z\}$ are the frequencies of the cavity mode, the $\sigma_\pm$ polarised lasers, and the magnetic splitting of the ground states, respectively, $\omega_{F'}$ are the frequencies of the excited hyperfine levels (we neglect magnetic shifts in these levels), $\Omega_\pm$ are the Rabi oscillation frequencies of the $\sigma_\pm$ polarised lasers and $g$ is the single-atom-cavity coupling constant (for the $^{87}$Rb $D_2$ line cycling transition). We then adiabatically eliminate the excited states following the method set out in Chapter 6 of \cite{GrimsmoThesis}. For this we rotate about $\hat{H}_0$, and then build two operators. The first of these, $\hat{Q}^\dagger$, is the sum of 
\begin{align}
\hat{Q}^\dagger_c &= \frac{g \mathrm{e}^{i\omega_ct}}{\sqrt{12}} \hat{a}^\dagger \left( \mathrm{e}^{-i\omega_zt}\ket{-1}\bra{1,-1} + \sqrt{3} \mathrm{e}^{-i\omega_zt}\ket{-1}\bra{2,-1} + 2\ket{0}\bra{2,0} - \mathrm{e}^{i\omega_zt}\ket{+1}\bra{1,+1} + \sqrt{3}\mathrm{e}^{i\omega_zt} \ket{+1}\bra{2,+1}\right), \\
\hat{Q}^\dagger_+ &= -\frac{\Omega_+\mathrm{e}^{i\omega_+t}}{2\sqrt{12}} \left(\mathrm{e}^{-i\omega_zt}\ket{-1}\bra{1,0} + \mathrm{e}^{-i\omega_zt}\ket{-1}\bra{2,0} + \ket{0}\bra{1,+1} + \sqrt{3} \ket{0} \bra{2,+1} + \sqrt{6} \mathrm{e}^{i\omega_zt}\ket{+1}\bra{2,+2}\right), \\
\hat{Q}^\dagger_- &= \frac{\Omega_-\mathrm{e}^{i\omega_-t}}{2\sqrt{12}} \left( - \sqrt{6}\mathrm{e}^{-i\omega_zt} \ket{-1}\bra{2,-2} + \ket{0} \bra{1,-1} - \sqrt{3} \ket{0} \bra{2,-1} + \mathrm{e}^{i\omega_zt}\ket{+1} \bra{1,0} - \mathrm{e}^{i\omega_zt}\ket{+1} \bra{2,0}\right),
\end{align}
\normalsize while the second, $\hat{\tilde{Q}}$, is the sum of
\begin{align}
\hat{\tilde{Q}}_c &= \frac{g\mathrm{e}^{-i\omega_ct}}{\sqrt{12}} \hat{a} \left( \frac{\mathrm{e}^{i\omega_zt}}{\Delta_c}\ket{1,-1}\bra{-1} + \frac{\sqrt{3}\mathrm{e}^{i\omega_zt}}{\Delta_c + \zeta} \ket{2,-1}\bra{-1} + \frac{2}{\Delta_c + \zeta} \ket{2,0} \bra{0} - \frac{\mathrm{e}^{-i\omega_zt}}{\Delta_c}\ket{1,+1}\bra{+1} + \frac{\sqrt{3}\mathrm{e}^{-i\omega_zt}}{\Delta_c + \zeta}\ket{2,+1}\bra{+1}\right), \\ 
\hat{\tilde{Q}}_+ &= -\frac{\Omega_+\mathrm{e}^{-i\omega_+t}}{2\sqrt{12}} \left( \frac{\mathrm{e}^{i\omega_zt}}{\Delta_+}\ket{1,0}\bra{-1} + \frac{\mathrm{e}^{i\omega_zt}}{\Delta_+ + \zeta}\ket{2,0}\bra{-1} + \frac{1}{\Delta_+} \ket{1,+1}\bra{0} +  \frac{\sqrt{3}}{\Delta_+ + \zeta}\ket{2,+1}\bra{0} + \frac{\sqrt{6}\mathrm{e}^{-i\omega_zt}}{\Delta_+ + \zeta} \ket{2,+2}\bra{+1}\right), \\
\hat{\tilde{Q}}_- &= \frac{\Omega_-\mathrm{e}^{-i\omega_-t}}{2\sqrt{12}} \left( - \frac{\sqrt{6}\mathrm{e}^{i\omega_zt}}{\Delta_- + \zeta} \ket{2,-2}\bra{-1} + \frac{1}{\Delta_-}\ket{1,-1}\bra{0} - \frac{\sqrt{3}}{\Delta_- + \zeta} \ket{2,-1}\bra{0} + \frac{\mathrm{e}^{-i\omega_zt}}{\Delta_-} \ket{1,0}\bra{+1} - \frac{\mathrm{e}^{-i\omega_zt}}{\Delta_-+\zeta}\ket{2,0} \bra{+1}\right),
\end{align}
\normalsize
where $\Delta_{c,\pm}$ are the detunings of the cavity mode and lasers from the $F'=1$ manifold (e.g., $\Delta_c = \omega_1 - \omega_c$) and $\zeta = \omega_2 - \omega_1$ is the frequency difference between the $F'=1$ and $F'=2$ manifolds (again neglecting any magnetic shifts). The effective Hamiltonian after the adiabatic elimination of the excited states is then 
\begin{equation}
\hat{H}_{\mathrm{eff}} = \hat{Q}^\dagger \hat{\tilde{Q}}.
\end{equation} 
This effective Hamiltonian (henceforth called $\hat{H}$) can then be written in the general form $\hat{H} = \sum\limits_{i,j = -1}^{+1} c_{i,j} \ket{i} \bra{j}$. We want to move this to a form in terms of angular momentum operators and nematic tensor terms. We first look at the terms of the Hamiltonian where $i=j$, with coefficients
\begin{align}
c_{-1,-1} &= \frac{g^2}{12\Delta_c} \hat{a}^\dagger \hat{a} + \frac{g^2}{4(\Delta_c + \zeta)} \hat{a}^\dagger \hat{a} + \frac{\Omega_+^2}{48\Delta_+} + \frac{\Omega_+^2}{48(\Delta_+ + \zeta)} + \frac{\Omega_-^2}{8(\Delta_-+\zeta)} ,\\
c_{0,0} &= \frac{g^2}{3(\Delta_c + \zeta)} \hat{a}^\dagger \hat{a} + \frac{\Omega_+^2}{48\Delta_+} + \frac{\Omega_+^2}{16(\Delta_++\zeta)} + \frac{\Omega_-^2}{48\Delta_-} + \frac{\Omega_-^2}{16(\Delta_-+\zeta)} ,\\
c_{+1,+1} &= \frac{g^2}{12\Delta_c} \hat{a}^\dagger \hat{a} + \frac{g^2}{4(\Delta_c + \zeta)} \hat{a}^\dagger \hat{a} + \frac{\Omega_+^2}{8(\Delta_+ + \zeta)} + \frac{\Omega_-^2}{48\Delta_-} + \frac{\Omega_-^2}{48(\Delta_- + \zeta)} .
\end{align}
We can use the spin-1 angular momentum operators $\hat{S}_z = \ket{+1}\bra{+1} - \ket{-1}\bra{-1}$ and $\hat{S}_z^2 = \ket{-1}\bra{-1} + \ket{+1}\bra{+1}$, and the identity $\mathbb{I} = \ket{-1}\bra{-1} + \ket{0}\bra{0} + \ket{+1}\bra{+1}$, to rewrite these terms as (where we discard a shift to the vacuum point)
\begin{equation}
\hat{H}_1 = \omega \hat{a}^\dagger \hat{a} + \omega_0 \hat{S}_z + \omega_q \hat{S}_z^2 + \frac{\delta_q}{2} \hat{S}_z^2 \hat{a}^\dagger \hat{a} ,
\end{equation}
where
\begin{align}
\omega &= \frac{g^2}{3(\Delta_c + \zeta)} ,\\
\omega_0 &= \frac{1}{96} \left( -\frac{5\Omega_+^2}{\Delta_+ + \zeta} + \frac{5\Omega_-^2}{\Delta_- + \zeta} + \frac{\Omega_+^2}{\Delta_+} - \frac{\Omega_-^2}{\Delta_-} \right),\\
\omega_q &= \frac{1}{96} \left( \frac{\Omega_+^2}{\Delta_++\zeta} + \frac{\Omega_-^2}{\Delta_- + \zeta} - \frac{\Omega_+^2}{\Delta_+} - \frac{\Omega_-^2}{\Delta_-} \right),\\
\delta_q &= \frac{g^2}{6} \left( \frac{1}{\Delta_c} - \frac{1}{\Delta_c + \zeta} \right).
\end{align}
We then consider terms where $i = j\pm1$, with coefficients
\begin{align}
c_{-1,0} &= \frac{g\Omega_-\mathrm{e}^{i(\omega_c - \omega_- - \omega_z)t}}{24} \left(\frac{1}{\Delta_-} - \frac{3}{\Delta_- + \zeta}\right) \hat{a}^\dagger - \frac{2g\Omega_+\mathrm{e}^{i(\omega_+ - \omega_c - \omega_z)t}}{24(\Delta_c + \zeta)} \hat{a} ,\\
c_{0,+1} &= \frac{g\Omega_+\mathrm{e}^{i(\omega_+-\omega_c-\omega_z)t}}{24} \left( \frac{1}{\Delta_c} - \frac{3}{\Delta_c + \zeta} \right) \hat{a} - \frac{2g\Omega_-\mathrm{e}^{i(\omega_c - \omega_- - \omega_z)t}}{24(\Delta_-+\zeta)} \hat{a}^\dagger ,\\
c_{0,-1} &= \frac{g\Omega_-\mathrm{e}^{i(\omega_- - \omega_c + \omega_z)t}}{24} \left( \frac{1}{\Delta_c} - \frac{3}{\Delta_c + \zeta} \right) \hat{a} - \frac{2g\Omega_+\mathrm{e}^{i(\omega_c - \omega_+ + \omega_z)t}}{24(\Delta_+ + \zeta)} \hat{a}^\dagger ,\\
c_{+1,0} &= \frac{g\Omega_+\mathrm{e}^{i(\omega_c - \omega_+ + \omega_z)t}}{24} \left( \frac{1}{\Delta_+} - \frac{3}{\Delta_+ + \zeta} \right) \hat{a}^\dagger - \frac{2g\Omega_-\mathrm{e}^{i(\omega_- - \omega_c + \omega_z)t}}{24(\Delta_c + \zeta)} \hat{a}.
\end{align}
To unravel this we use $\hat{S}_+  = \sqrt{2} ( \ket{+1} \bra{0} + \ket{0} \bra{-1} )$ and $\hat{S}_- = \sqrt{2} ( \ket{-1} \bra{0} + \ket{0} \bra{+1} )$, as well as the nematic tensor terms $\hat{Q}_{xz} = \sqrt{2} (\ket{+1}\bra{0} + \ket{0}\bra{+1} - \ket{0} \bra{-1} - \ket{-1} \bra{0})$ and $i\hat{Q}_{yz} = \sqrt{2} ( \ket{+1} \bra{0} - \ket{0} \bra{+1} - \ket{0} \bra{-1} + \ket{-1} \bra{0})$. This gives us
\begin{equation}
\hat{H}_2 = \frac{\lambda_1}{\sqrt{2}} \hat{S}_+ \hat{a} + + \frac{\lambda_2}{\sqrt{2}} \hat{S}_- \hat{a}^\dagger + \frac{\lambda_3}{\sqrt{2}} \hat{S}_- \hat{a} + \frac{\lambda_4}{\sqrt{2}} \hat{S}_+ \hat{a}^\dagger + \frac{\xi_1}{\sqrt{2}} \hat{Q}_{xz} \hat{a} + \frac{\xi_2}{\sqrt{2}} \hat{Q}_{xz} \hat{a}^\dagger + \frac{i\xi_3}{\sqrt{2}} \hat{Q}_{yz} \hat{a} + \frac{i\xi_4}{\sqrt{2}} \hat{Q}_{yz} \hat{a}^\dagger ,
\end{equation}
where the parameters are
\begin{align}
\lambda_1 &= \frac{g\Omega_-\mathrm{e}^{i(\omega_- - \omega_c + \omega_z)t}}{48} \left( \frac{1}{\Delta_c} - \frac{5}{\Delta_c + \zeta} \right) ,\\
\lambda_2 &= \frac{g\Omega_-\mathrm{e}^{i(\omega_c - \omega_- - \omega_z)t}}{48} \left( \frac{1}{\Delta_-} - \frac{5}{\Delta_- + \zeta} \right) ,\\
\lambda_3 &= \frac{g\Omega_+\mathrm{e}^{i(\omega_+ - \omega_c - \omega_z)t}}{48} \left( \frac{1}{\Delta_c} - \frac{5}{\Delta_c + \zeta} \right) ,\\
\lambda_4 &= \frac{g\Omega_+\mathrm{e}^{i(\omega_c - \omega_+ + \omega_z)t}}{48} \left( \frac{1}{\Delta_+} - \frac{5}{\Delta_+ + \zeta} \right) ,\\
\xi_1 &= \frac{g\Omega_-\mathrm{e}^{i(\omega_- - \omega_c + \omega_z)t}}{96} \left( \frac{1}{\Delta_c + \zeta} - \frac{1}{\Delta_c} \right) + \frac{g\Omega_+\mathrm{e}^{i(\omega_+ - \omega_c - \omega_z)t}}{96} \left( \frac{1}{\Delta_c} - \frac{1}{\Delta_c + \zeta} \right) ,\\
\xi_2 &=  \frac{g\Omega_+\mathrm{e}^{i(\omega_c - \omega_+ + \omega_z)t}}{96} \left( \frac{1}{\Delta_+} - \frac{1}{\Delta_+ + \zeta} \right) + \frac{g\Omega_-\mathrm{e}^{i(\omega_c - \omega_- - \omega_z)t}}{96} \left( \frac{1}{\Delta_- + \zeta} - \frac{1}{\Delta_-}\right) ,\\
\xi_3 &= \frac{g\Omega_-\mathrm{e}^{i(\omega_- - \omega_c + \omega_z)t}}{96} \left( \frac{1}{\Delta_c + \zeta} - \frac{1}{\Delta_c} \right) + \frac{g\Omega_+\mathrm{e}^{i(\omega_+ - \omega_c - \omega_z)t}}{96} \left( \frac{1}{\Delta_c + \zeta} - \frac{1}{\Delta_c} \right) ,\\
\xi_4 &= \frac{g\Omega_+\mathrm{e}^{i(\omega_c - \omega_+ + \omega_z)t}}{96} \left( \frac{1}{\Delta_+} - \frac{1}{\Delta_+ + \zeta} \right) + \frac{g\Omega_-\mathrm{e}^{i(\omega_c - \omega_- - \omega_z)t}}{96} \left( \frac{1}{\Delta_-} - \frac{1}{\Delta_- + \zeta} \right) .
\end{align}
We now consider the two terms that involve a complete spin flip (i.e., $i = -j$) and note that, using $\hat{S}_-^2 = 2\ket{-1}\bra{+1}$ and $\hat{S}_+^2 = 2\ket{+1}\bra{-1}$, this can be immediately rewritten as
\begin{equation}
\hat{H}_3 = h_+ \hat{S}_+^2 + h_- \hat{S}_-^2 ,
\end{equation}
where
\begin{align}
h_+ &= \frac{\Omega_+\Omega_-\mathrm{e}^{i(\omega_- - \omega_+ + 2\omega_z)t}}{96} \left( \frac{1}{\Delta_+} - \frac{1}{\Delta_+ + \zeta} \right) ,\\
h_- &= \frac{\Omega_+\Omega_-\mathrm{e}^{i(\omega_+ - \omega_- - 2\omega_z)t}}{96} \left( \frac{1}{\Delta_-} - \frac{1}{\Delta_- + \zeta} \right).
\end{align}
All parts of the Hamiltonian can be made time-independent by rotating about
\begin{equation}
\hat{H}_r = \left(-\omega_c + \frac{\omega_+ + \omega_-}{2}\right) \hat{a}^\dagger \hat{a} + \left( - \omega_z + \frac{\omega_+ - \omega_-}{2} \right) \hat{S}_z ,
\end{equation}
which removes all time dependence from the parameters and changes a few such that
\begin{align}
\omega &= \frac{g^2}{3(\Delta_c + \zeta)} + \omega_c - \frac{\omega_+ + \omega_-}{2} ,\\
\omega_0 &= \frac{1}{96} \left( -\frac{5\Omega_+^2}{\Delta_+ + \zeta} + \frac{5\Omega_-^2}{\Delta_- + \zeta} + \frac{\Omega_+^2}{\Delta_+} - \frac{\Omega_--^2}{\Delta_-} \right) + \omega_z + \frac{\omega_- - \omega_+}{2}.
\end{align}
We then make the assumption of very large detuning, such that $\Delta_-^{-1} \approx \Delta_+^{-1} \approx \Delta_c^{-1}$, and so substitute a generic $\Delta$ for all of these detunings. We can then write our full, effective Hamiltonian as
\begin{align}
\hat{H} = \omega \hat{a}^\dagger \hat{a} + \omega_0 \hat{S}_z + \omega_q \hat{S}_z^2 + \frac{\delta_q}{2} \hat{S}_z^2 \hat{a}^\dagger \hat{a} + \frac{\lambda_1}{\sqrt{2}}(\hat{S}_+ \hat{a} + \hat{S}_- \hat{a}^\dagger) + \frac{\lambda_2}{\sqrt{2}} (\hat{S}_- \hat{a} + \hat{S}_+ \hat{a}^\dagger) \nonumber\\+ \xi_1 \hat{Q}_{xz} (\hat{a} + \hat{a}^\dagger) + i \xi_2 \hat{Q}_{yz} (\hat{a} - \hat{a}^\dagger) + h (\hat{S}_+^2 + \hat{S}_-^2)
\end{align}
\normalsize
with parameters
\begin{align}
\omega &= \frac{g^2}{3(\Delta + \zeta)} + \omega_c - \frac{\omega_+ + \omega_-}{2} ,\\
\omega_0 &= \frac{\Omega_-^2 - \Omega_+^2}{96} \left( \frac{5}{\Delta + \zeta}  - \frac{1}{\Delta} \right) + \omega_z + \frac{\omega_- - \omega_+}{2} ,\\
\omega_q &= \frac{\Omega_+^2 + \Omega_-^2}{96} \left( \frac{1}{\Delta+\zeta} - \frac{1}{\Delta} \right)  ,\\
\delta_q &= \frac{g^2}{6} \left( \frac{1}{\Delta} - \frac{1}{\Delta + \zeta} \right) ,\\
\lambda_1 &= \frac{g\Omega_-}{48} \left( \frac{1}{\Delta} - \frac{5}{\Delta + \zeta} \right), \\
\lambda_2 &= \frac{g\Omega_+}{48} \left( \frac{1}{\Delta} - \frac{5}{\Delta + \zeta} \right) ,\\
\xi_1 &= \frac{g\Omega_-}{96} \left(\frac{1}{\Delta + \zeta} - \frac{1}{\Delta}\right) + \frac{g\Omega_+}{96} \left( \frac{1}{\Delta} - \frac{1}{\Delta + \zeta} \right) ,\\
\xi_2 &= \frac{g\Omega_-}{96} \left( \frac{1}{\Delta + \zeta} - \frac{1}{\Delta}\right) + \frac{g\Omega_+}{96} \left( \frac{1}{\Delta + \zeta} - \frac{1}{\Delta} \right) ,\\
h &= \frac{\Omega_+\Omega_-}{96} \left( \frac{1}{\Delta} - \frac{1}{\Delta + \zeta} \right).
\end{align}
With $\Delta \gg \zeta$, we find that all of the parameters $\{\omega_q, \delta_q, \xi_1, \xi_2, h \}$ are negligible, which leaves the simple Hamiltonian
\begin{equation}
\hat{H} = \omega \hat{a}^\dagger \hat{a} + \omega_0 \hat{S}_z + \frac{\lambda_1}{\sqrt{2}} (\hat{S}_+ \hat{a} + \hat{S}_- \hat{a}^\dagger) + \frac{\lambda_2}{\sqrt{2}} (\hat{S}_- \hat{a} + \hat{S}_+ \hat{a}^\dagger) .
\end{equation}
Summing over an ensemble of $N$ atoms yields
\begin{align}
\hat{H} &= \omega \hat{a}^\dagger \hat{a} + \sum\limits_{i = 1}^N \left\{ \omega_0 \hat{S}_z^{(i)} + \frac{\lambda_1}{\sqrt{2}} (\hat{S}_+^{(i)} \hat{a} + \hat{S}_-^{(i)} \hat{a}^\dagger) + \frac{\lambda_2}{\sqrt{2}} (\hat{S}_-^{(i)} \hat{a} + \hat{S}_+^{(i)} \hat{a}^\dagger) \right\} \\
&= \omega \hat{a}^\dagger \hat{a} + \omega_{0} \hat{\tilde{S}}_z + \frac{\lambda_-}{\sqrt{2N}} (\hat{\tilde{S}}_+ \hat{a} + \hat{\tilde{S}}_- \hat{a}^\dagger) + \frac{\lambda_+}{\sqrt{2N}} (\hat{\tilde{S}}_- \hat{a} + \hat{\tilde{S}}_+ \hat{a}^\dagger)
\end{align}
where we have effective parameters
\begin{align}
\omega &= \frac{Ng^2}{3\Delta} + \omega_c - \frac{\omega_+ + \omega_-}{2} , \\
\omega_0 &= \frac{\Omega_-^2 - \Omega_+^2}{24 \Delta} + \omega_z + \frac{\omega_- - \omega_+}{2}, \\
\lambda_- &= - \frac{\sqrt{N}g\Omega_-}{12\Delta} ,\\
\lambda_+ &= - \frac{\sqrt{N}g\Omega_+}{12\Delta},
\end{align}
and we have defined collective spin operators
\begin{align}
\hat{\tilde{S}}_z = \sum\limits_{i=1}^N \hat{S}_z^{(i)} , ~~~~
\hat{\tilde{S}}_\pm = \sum\limits_{i=1}^N \hat{S}_\pm^{(i)}.
\end{align}

\section{Adiabatic elimination of the cavity mode}

We now consider the Hamiltonian in the limit that $\sqrt{\omega^2 + \kappa^2} \gg |\omega_0|, |\lambda_{r,s}|$ (note that whilst we do consider $\omega \gg \kappa$ in our simulations, this is not necessary for the derivation). We split our Hamiltonian into two parts, $\hat{H} = \hat{H}_{0} + \hat{H}_{i}$, where
\begin{align}
\hat{H}_0 &= \omega \hat{a}^\dagger \hat{a} + \omega_0 \hat{S}_z ,\\
\hat{H}_i &=  \frac{\lambda_-}{\sqrt{2j}} (\hat{S}_+ \hat{a} + \hat{S}_- \hat{a}^\dagger) + \frac{\lambda_+}{\sqrt{2j}} (\hat{S}_- \hat{a} + \hat{S}_+ \hat{a}^\dagger) ,
\end{align}
and move to the interaction picture defined by
\begin{equation}
\tilde{\rho} (t) = \mathrm{e}^{i\hat{H}_0t}\rho(t)\mathrm{e}^{-i\hat{H}_0t} .
\end{equation}
It follows that
\begin{equation}
\frac{\partial \tilde{\rho}}{\partial t} = -i[\hat{\tilde{H}}_i,\tilde{\rho}] + \mathcal{L}_c \tilde{\rho}(t)
\end{equation}
where
\begin{equation}
\mathcal{L}_c \rho \equiv \kappa \left(2 \hat{a} \rho \hat{a}^\dagger - \hat{a}^\dagger \hat{a} \rho - \rho \hat{a}^\dagger \hat{a} \right) ,
\end{equation}
and our interaction Hamiltonian is
\begin{align}
\hat{\tilde{H}}_i &= \mathrm{e}^{i\hat{H}_0t} H_{i} \mathrm{e}^{-i\hat{H}_0t} \nonumber \\
&= \frac{1}{\sqrt{2j}} \left[ \hat{a} \mathrm{e}^{-i\omega t} (\lambda_- \hat{S}_+ \mathrm{e}^{i\omega_0t} + \lambda_+ \hat{S}_- \mathrm{e}^{-i\omega_0t}) + \hat{a}^\dagger \mathrm{e}^{i\omega t} (\lambda_- \hat{S}_- \mathrm{e}^{-i\omega_0t} + \lambda_+ \hat{S}_+ \mathrm{e}^{i\omega_0t}) \right] \nonumber \\
&= \hat{a} \mathrm{e}^{-i\omega t} \hat{X}(t) + \hat{a}^\dagger \mathrm{e}^{i\omega t} \hat{X}^\dagger (t) ,
\end{align}
where
\begin{equation}
\hat{X}(t) = \frac{\lambda_-}{\sqrt{2N}} \hat{S}_- \mathrm{e}^{-i\omega_0t} + \frac{\lambda_+}{\sqrt{2N}} \hat{S}_+ \mathrm{e}^{i\omega_0t}.
\end{equation}
We want $\tilde{\rho}_s = \mathrm{Tr}_c\{ {\tilde{\rho}}\}$, and so, in the second-order Born-Markov approximation, we have that
\begin{equation}\label{eq:ME2ndOrder}
\frac{\partial \tilde{\rho}_s}{\partial t} = - \int\limits_0^t \mathrm{d}t' \mathrm{Tr}_c \left\{ \left[\hat{\tilde{H}}_i , \mathrm{e}^{\mathcal{L}_c (t-t')} [ \hat{\tilde{H_i}}(t'), \tilde{\rho}_s(t') \otimes \tilde{\rho}_c(t') ] \right] \right\}.
\end{equation}
Expanding this out, using the cyclical property of the trace, and assuming that the cavity mode is essentially in the vacuum state, so that the only non-zero traces are
\begin{align}
&\mathrm{Tr}_c \{ \hat{a} \mathrm{e}^{\mathcal{L}_c(t-t')} \hat{a}^\dagger \tilde{\rho}_c (t') \} = \braket{ \hat{a}(t) \hat{a}^\dagger (t') } = \mathrm{e}^{-\kappa (t-t')} ,\\
&\mathrm{Tr}_c \{ \hat{a}^\dagger \mathrm{e}^{\mathcal{L}_c(t-t')} \tilde{\rho}_c (t') \hat{a}  \} = \braket{ \hat{a}(t') \hat{a}^\dagger (t) } = \mathrm{e}^{-\kappa (t-t')} ,
\end{align}
we can simplify the right-hand side of (\ref{eq:ME2ndOrder}) considerably. We further assume that in the integral we can set $\hat{X}(t') \approx \hat{X}(t)$ and $\tilde{\rho}_s(t') \approx \tilde{\rho}_s (t)$, so that it only remains to evaluate
\begin{equation}
\int\limits_0^t \mathrm{d} t' \mathrm{e}^{-(\kappa + i\omega)(t-t')} = \frac{1}{\kappa + i\omega} \left( 1- \mathrm{e}^{-(\kappa + i\omega)t} \right) \rightarrow \frac{1}{\kappa + i\omega} ,
\end{equation}
where we  assume that $t$ is sufficiently large to neglect the exponential. Thus, we obtain the master equation
\begin{align}
\frac{\partial \tilde{\rho}_s}{\partial t} =  &- \frac{1}{\kappa + i\omega} \hat{X}^\dagger(t) \hat{X} (t) \tilde{\rho}_s (t) + \frac{1}{\kappa + i\omega} \hat{X} (t) \tilde{\rho}_s (t) \hat{X}^\dagger(t) +\frac{1}{\kappa - i\omega} \hat{X} (t) \tilde{\rho}_s (t)\hat{X}^\dagger(t) - \frac{1}{\kappa - i\omega} \tilde{\rho}_s (t) \hat{X}^\dagger(t) \hat{X}(t) 
\nonumber
\\
&= \frac{2\kappa}{\kappa^2 + \omega^2} \hat{X} (t) \tilde{\rho}_s (t) \hat{X}^\dagger(t) - \frac{\kappa - i\omega}{\kappa^2+\omega^2} \hat{X}^\dagger(t) \hat{X}(t) \tilde{\rho}_s(t) - \frac{\kappa + i\omega}{\kappa^2+\omega^2} \tilde{\rho}_s(t) \hat{X}^\dagger(t)\hat{X}(t) \nonumber \\
&= i [\frac{\omega}{\kappa^2 + \omega^2} \hat{X}^\dagger(t) \hat{X}(t), \tilde{\rho}_s(t) ] + \frac{\kappa}{\kappa^2 + \omega^2} \left( 2\hat{X} \tilde{\rho}_s(t) \hat{X}^\dagger(t) - \hat{X}^\dagger(t) \hat{X} (t) \tilde{\rho}_s(t) - \tilde{\rho}_s (t)\hat{X}^\dagger(t)\hat{X}(t) \right).
\end{align}
We rotate back out of the interaction picture to give
\begin{align}
\frac{\partial \rho_s}{\partial t} &= -i[\hat{H},\rho_s] + \frac{\kappa}{\kappa^2 + \omega^2} (2 \hat{X} \rho_s \hat{X}^\dagger - \hat{X}^\dagger \hat{X} \rho_s - \rho_s \hat{X}^\dagger \hat{X}) ,
\end{align}
where
\begin{align}
\hat{H} = \omega_0 \hat{S}_z - \frac{\omega}{\kappa^2 + \omega^2} \hat{X}^\dagger \hat{X} , ~~~~~~\hat{X} = \frac{\lambda_-}{\sqrt{2N}} \hat{S}_- + \frac{\lambda_+}{\sqrt{2N}} \hat{S}_+ .
\end{align}
We can substitute back in for $\hat{X}$ and transform the Hamiltonian to
\begin{align}
\hat{H} &= \omega_0 \hat{S}_z - \frac{\omega}{2N(\kappa^2 + \omega^2)}(\lambda_- \hat{S}_+ + \lambda_+ \hat{S}_-)(\lambda_- \hat{S}_- + \lambda_+ \hat{S}_+)  \nonumber \\
&= \omega_0 \hat{S}_z - \frac{\omega}{2N(\kappa^2 + \omega^2)} \left[ (\lambda_- + \lambda_+)^2 \hat{S}_x^2 + (\lambda_- - \lambda_+)^2 \hat{S}_y^2 + (\lambda_-^2 - \lambda_+^2) \hat{S}_z \right]
\end{align} 
where we assume $\lambda_\pm$ are real.

\section{Simulating the system}

\subsection{Numerical simulation}

For the generation of spin-nematic squeezing, we want to simulate the master equation
\begin{equation}
\frac{\partial \rho}{\partial t} = -i\left[\frac{\Lambda}{2N} (\hat{S}_x^2 + \hat{S}_y^2), \rho\right] + \frac{\Gamma}{2N} (2\hat{S}_- \rho \hat{S}_+ - \hat{S}_+\hat{S}_- \rho - \rho \hat{S}_+\hat{S}_-).
\end{equation}
For the purposes of simulation we move to a basis of modal populations. We define $\hat{a}_m (\hat{a}_m^\dagger)$ as a bosonic annihilation (creation) operator for the spin state $m$. We can then identify $\hat{S}_- = \sqrt{2} (\hat{a}^\dagger_{-1} \hat{a}_0 + \hat{a}_0^\dagger \hat{a}_{+1})$ and $\hat{S}_+ = \sqrt{2} (\hat{a}^\dagger_0 \hat{a}_{-1} + \hat{a}^\dagger_{+1} \hat{a}_0)$ and rewrite the Hamiltonian in the form
\begin{equation}
\hat{H} = \frac{\Lambda}{2N} \left( 2\hat{a}^\dagger_{+1}\hat{a}^\dagger_{-1}\hat{a}_0\hat{a}_0 + 2\hat{a}_0^\dagger \hat{a}_0^\dagger \hat{a}_{+1} \hat{a}_{-1} + \hat{a}_0^\dagger \hat{a}_0 (1 + 2\hat{a}_{+1}^\dagger \hat{a}_{+1} + 2\hat{a}_{-1}^\dagger \hat{a}_{-1}) \right).
\end{equation}
We start our simulations with an initial state in which the entire population is in the $m=0$ state and use a Monte-Carlo wavefunction method to simulate the system. An average over an ensemble of trajectories is used to estimate the master equation result for the various moments required to calculate $\xi_x^2$.

\subsection{Undepleted mode approximation}

To obtain an analytic result we can make the approximation that the $m=0$ mode remains essentially undepleted, or, in other words, that $N\rightarrow\infty$. This enables us to make the replacement $\hat a_0 = \hat a_0^\dagger = \sqrt{N}$, so that the Hamiltonian becomes
\begin{equation}
\hat{H} = \frac{\Lambda}{2} \left( 2\hat{a}_{+1}^\dagger \hat{a}_{-1}^\dagger + 2\hat{a}_{+1}\hat{a}_{-1} + 1 + 2\hat{a}_{+1}^\dagger \hat{a}_{+1} + 2\hat{a}_{-1}^\dagger \hat{a}_{-1} \right) ,
\end{equation}
while in the damping term $\hat S_- \rightarrow \sqrt{2N} (\hat a_{-1}^\dag +\hat a_{+1})$ and $\hat S_+ \rightarrow \sqrt{2N} (\hat a_{-1} +\hat a_{+1}^\dag )$. 

Equations of motion for the various moments required to compute $\xi_x^2$ can be deduced from the master equation, but we choose to work instead with the equivalent quantum Langevin equations for $\hat{a}_{\pm1}$, which, defining a vacuum noise operator $\hat{b}_{\mathrm{in}}(t)$ that satisfies $[\hat{b}_{\mathrm{in}}(t),\hat{b}^\dagger_{\mathrm{in}}(t')] = \delta(t-t')$, take the simple linear forms
\begin{align}
\frac{\mathrm{d}\hat{a}_{+1}}{\mathrm{d}t} &= - (\Gamma + i\Lambda) (\hat{a}_{+1} + \hat{a}^\dagger_{-1}) - \sqrt{2\Gamma}\hat{b}_{\mathrm{in}}(t) ,\\
\frac{\mathrm{d}\hat{a}_{-1}}{\mathrm{d}t} &=  (\Gamma - i\Lambda) (\hat{a}_{-1} + \hat{a}^\dagger_{+1}) + \sqrt{2\Gamma}\hat{b}^\dagger_{\mathrm{in}}(t).
\end{align}
To calculate the squeezing parameter $\xi_x^2$ we require two combinations of these: $\hat{A} = \hat{a}_{+1} + \hat{a}_{-1}^\dagger$ and $\hat{B} = \hat{a}_{+1} -\hat{a}_{-1}^\dagger$. These have Langevin equations
\begin{align}
\frac{\mathrm{d}\hat{A}}{\mathrm{d}t} &= 0 ,\\
\frac{\mathrm{d}\hat{B}}{\mathrm{d}t} &= -2(\Gamma + i\Lambda) \hat{A} - 2\sqrt{2\Gamma}\hat{b}_{\mathrm{in}}(t) ,
\end{align}
which are readily integrated to give
\begin{align}
\hat{A}(t) &= \hat{A}(0) \\
\hat{B}(t) &= \hat{B}(0) - 2(\Gamma + i\Lambda) \hat{A}(0) t - 2\sqrt{2\Gamma}\int\limits_0^t \mathrm{d}t' \hat{b}_{\mathrm{in}}(t').
\end{align}
The specific operators of interest are given by
\begin{align}
\hat{S}_x(t) &= \sqrt{\frac{N}{2}} (\hat{A}(t) + \hat{A}^\dagger(t)) \\
\hat{Q}_{yz}(t) &=  i\sqrt{\frac{N}{2}} (\hat{B}(t) - \hat{B}^\dagger(t))\\
\end{align}
and using the results for $\hat{A}(t)$ and $\hat{B}(t)$ (and assuming that we can set $\hat{Q}_{zz} - \hat{Q}_{yy} = -2N$) we obtain
\begin{equation}
\xi_x^2 = (\cos\theta + 2\Lambda t\, \sin\theta )^2 + (1+2\Gamma t)^2\sin^2 \theta . \label{nocorrection}
\end{equation}
In the Letter, we compare this analytic result to the finite-$N$ numerical results. The formula above correctly predicts the angular dependence, and, at shorter times, also matches quite accurately the degree of squeezing for the $N=120$ case. 
For larger $N$ it should provide a good model of the squeezing for longer times, where it predicts that the degree of squeezing will continue to increase for a suitable choice of phase angle $\theta$. Note also that the result (\ref{nocorrection}) is unchanged if a term $\omega_0^\prime \hat{S}_z\equiv \omega_0^\prime (\hat{a}_{+1}^\dag \hat{a}_{+1}-\hat{a}_{-1}^\dag \hat{a}_{-1})$ is added to the Hamiltonian.

Below we compare the analytic results from (\ref{nocorrection}) for different levels of damping. It is clear that an increase in damping does reduce the level of squeezing at a given time, but that even for relatively large values of $\Gamma / \Lambda$, strong squeezing is still possible.

\begin{figure}[h]
\begin{subfigure}{0.32\textwidth}
\includegraphics[width=\textwidth]{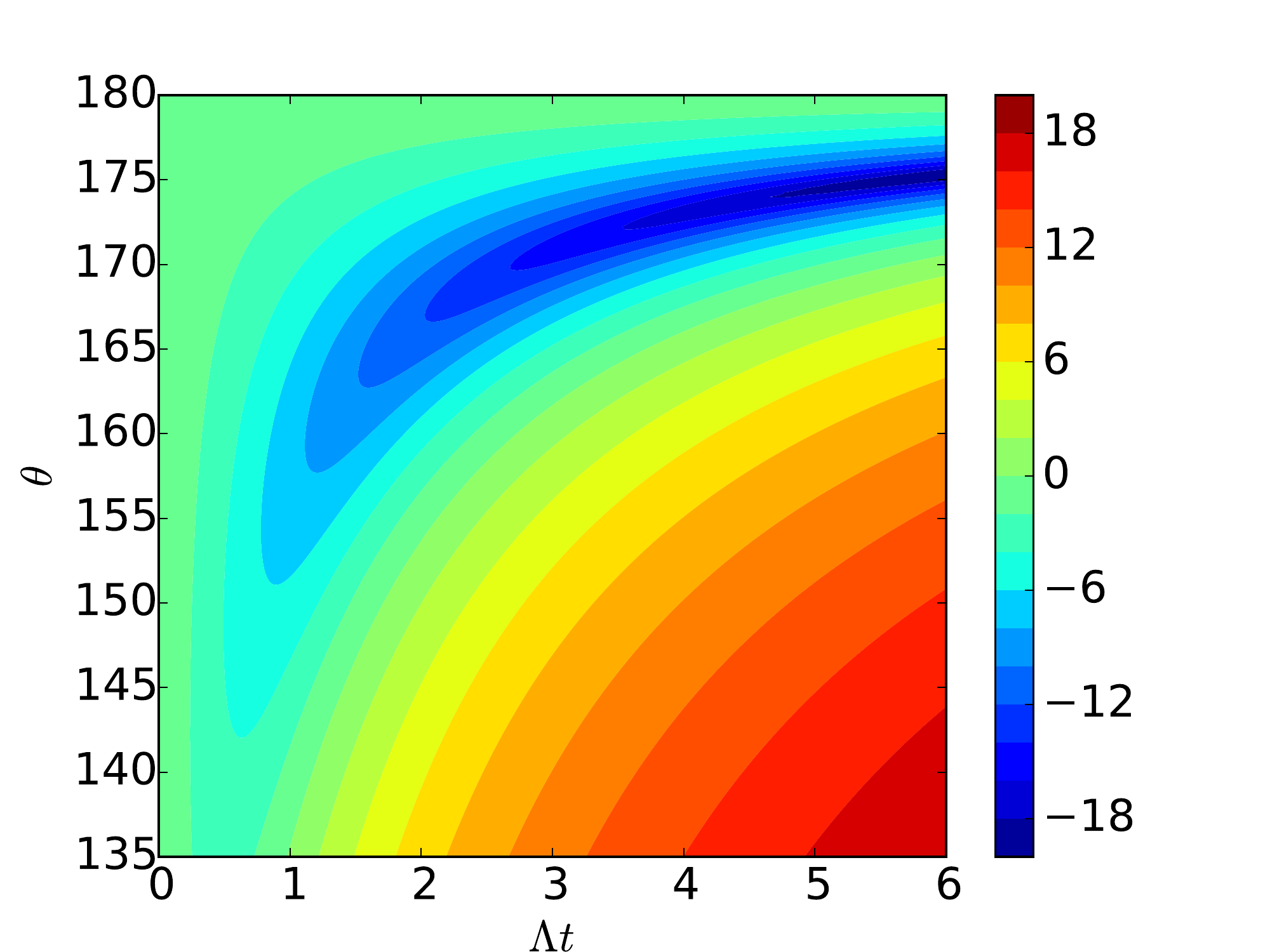}
\caption{$\Gamma / \Lambda = 0.02$}
\end{subfigure}
\begin{subfigure}{0.32\textwidth}
\includegraphics[width=\textwidth]{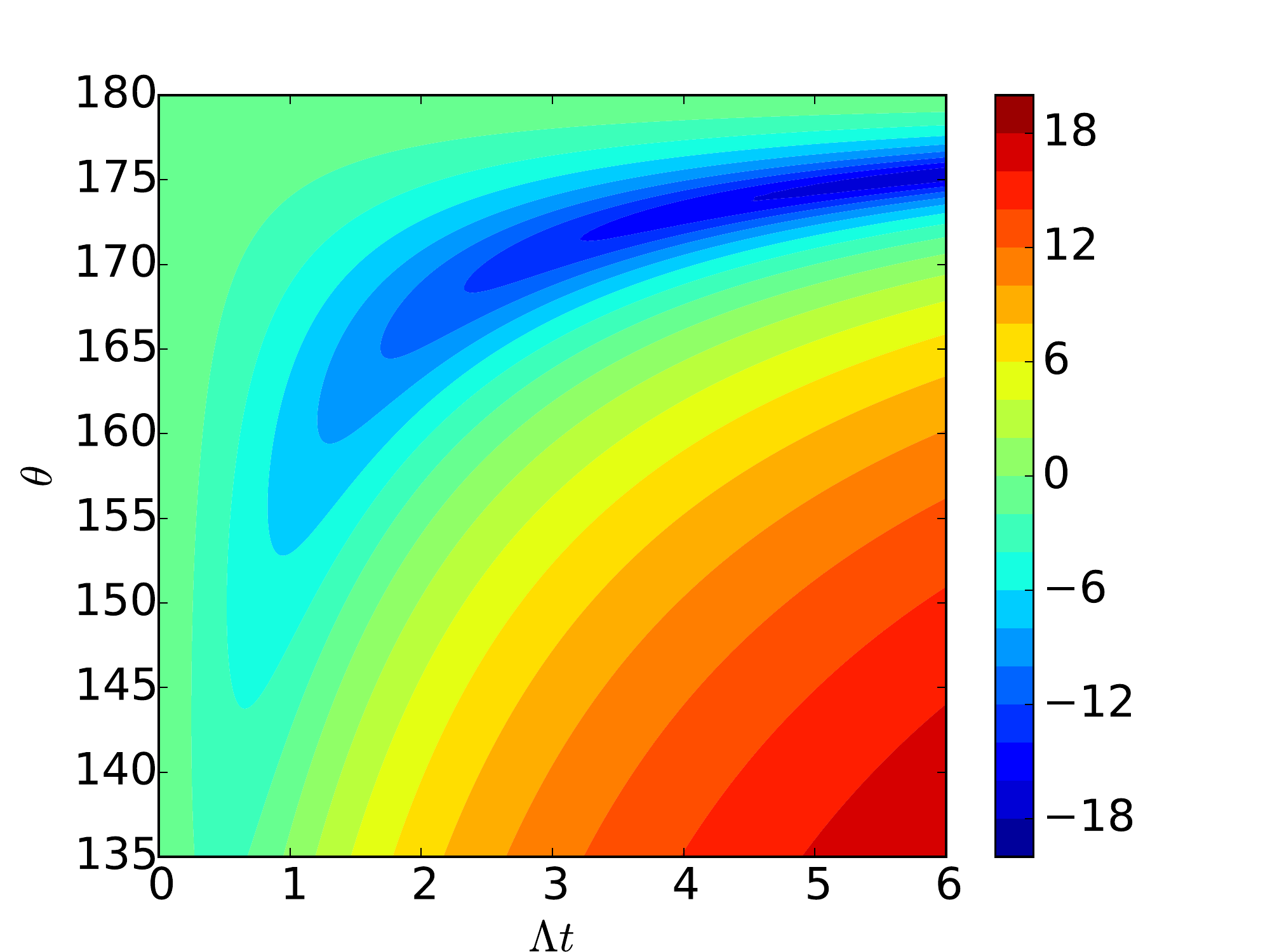}
\caption{$\Gamma / \Lambda = 0.05$}
\end{subfigure}
\begin{subfigure}{0.32\textwidth}
\includegraphics[width=\textwidth]{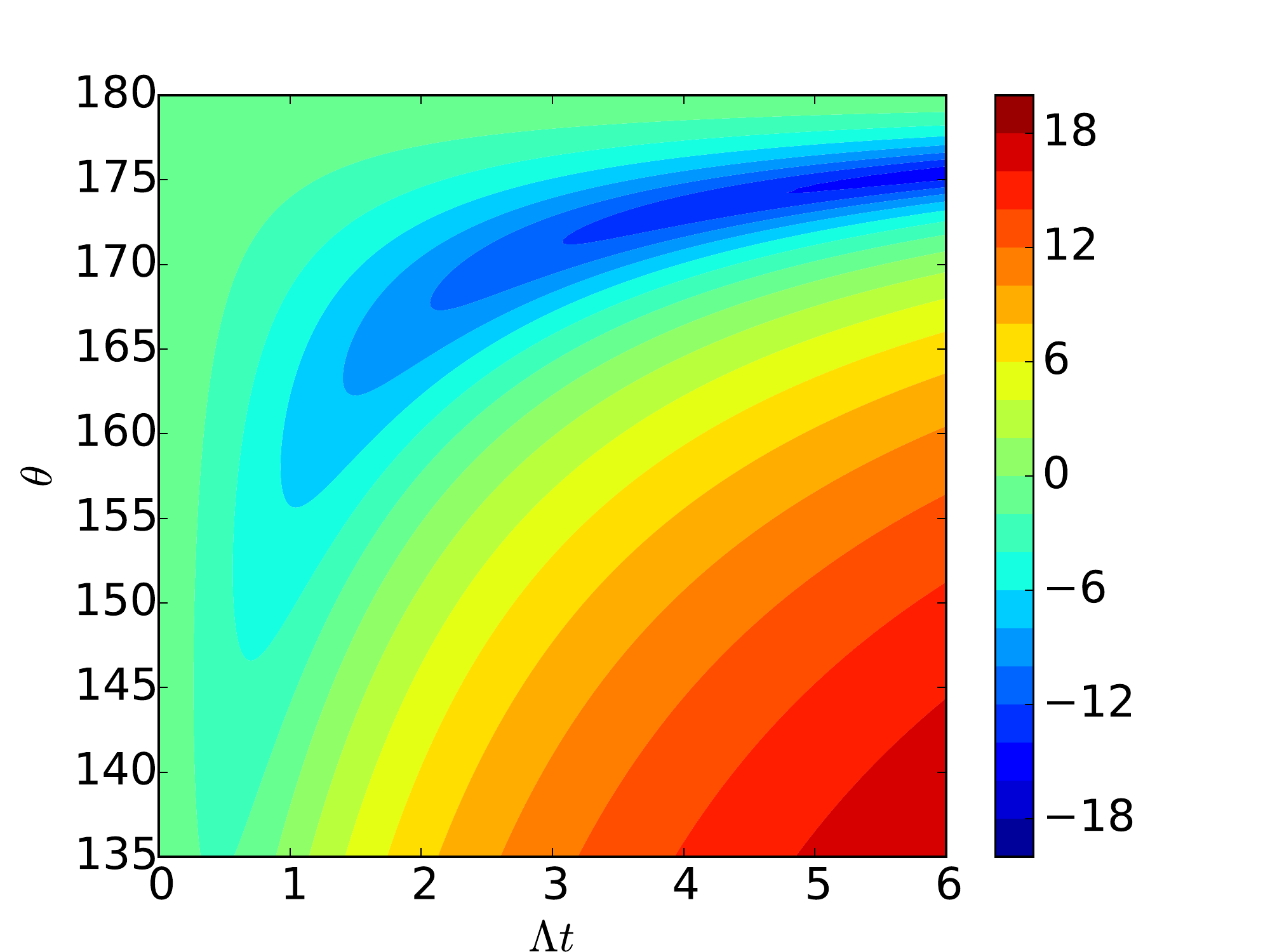}
\caption{$\Gamma / \Lambda = 0.1$}
\end{subfigure}
\caption{Values for $\xi_x^2 (\theta,t)$ (in dB) as a function of time and phase angle using Eq.~(\ref{nocorrection}).}
\end{figure}

\section{Visualising the atomic state}

\subsection{Defining coherent states and the Q-function for the space $\set{\hat{S}_x,\hat{Q}_{yz},\hat{Q}_{zz}-\hat{Q}_{yy}}$}

In the Letter, we describe squeezing that takes place on a sphere with axes $\set{\hat{S}_x,\hat{Q}_{yz},\hat{Q}_{zz}-\hat{Q}_{yy}}$ (or equivalently $\set{\hat{S}_y,\hat{Q}_{xz},\hat{Q}_{xx}-\hat{Q}_{zz}}$, with an equivalent derivation to here). With spin squeezing, it is useful to be able to visualise the state with the atomic Q-function. Analogously to the Q-function for electromagnetic modes, an overlap is taken with coherent states over the space of interest. Since we consider squeezing on the SU(2) sphere $\set{\hat{S}_x,\hat{Q}_{yz},\hat{Q}_{zz}-\hat{Q}_{yy}}$, we build coherent states of that space. We build these in the same way as coherent spin states of spin-1/2 particles: a binomial distribution across the eigenstates of the $z$ axis \cite{Arecchi72,Gross12}, which in this case is $\hat{Q}_{zz} - \hat{Q}_{yy}$. In terms of the bosonic annihilation and creation operators, this is
\begin{equation}
\hat{Q}_{zz} - \hat{Q}_{yy} = -2 \hat{a}^\dagger_0\hat{a}_0 + \hat{a}^\dagger_{+1} \hat{a}_{+1} + \hat{a}^\dagger_{-1}\hat{a}_{-1} + \hat{a}^\dagger_{+1} \hat{a}_{-1} + \hat{a}^\dagger_{-1} \hat{a}_{+1}
\end{equation}
It is easy to see that the initial state $\ket{0,N,0}$ is an eigenstate of this operator with eigenvalue $-2N$. We define this lowest energy eigenstate as the South pole of our sphere and notate it as $\ket{0}$. To generate the other eigenstates we use the ladder operators
\begin{equation}
\hat{\cal S}_\pm = \hat{S}_x \pm i\hat{Q}_{yz} = 2\sqrt{2} \hat{a}_0 (\hat{a}^\dagger_{+1} + \hat{a}^\dagger_{-1}).
\end{equation}
Applying this operator to our lowest energy eigenstates gives us the rest of the eigenstates,
\begin{equation}
\ket{M} = (\hat{\cal S}_+)^M \ket{0},
\end{equation}
where we have omitted a normalisation factor. We then build coherent states as
\begin{equation}
\ket{\eta = \mathrm{e}^{i\phi}\mathrm{tan}(\theta/2)} = \sum\limits^{N}_{M=0} \left( \begin{matrix} N \\ M \end{matrix} \right)^{1/2} \eta^M \ket{M}.
\end{equation}
Our atomic Q-function is described by
\begin{equation}
Q(\eta) = \bra{\eta} \rho \ket{\eta}
\end{equation}
and $(\theta,\phi)$ are mapped as the polar and azimuthal angles onto the surface of a sphere.

\subsection{Q-function plots of the state}

As discussed in the Letter, our squeezing generator is analogous to that in four-wave mixing. Correlated pairs of atoms in the $m=\pm1$ states are created from a reservoir of atoms in the $m=0$ state. Unlike in the optical analogy, here the reservoir has a finite number of atoms in it, and so the squeezing takes place on a sphere instead of a plane. As pairs are created, $|\braket{\hat{Q}_{zz}-\hat{Q}_{yy}}|$ begins to reduce. In terms of the atomic Q-function plots this can be seen as the ends of the state travelling up the sphere. At some point this effect reduces $|\braket{\hat{Q}_{zz}-\hat{Q}_{yy}}|$ at a faster rate than the squeezing generator is reducing the variance, and so the squeezing parameter begins to increase. We also see a slight ``twist'' developing in the state where the extreme ends are at slightly exaggerated angles.

\begin{figure}[h]
\begin{subfigure}{0.3\textwidth}
\includegraphics[width=\textwidth]{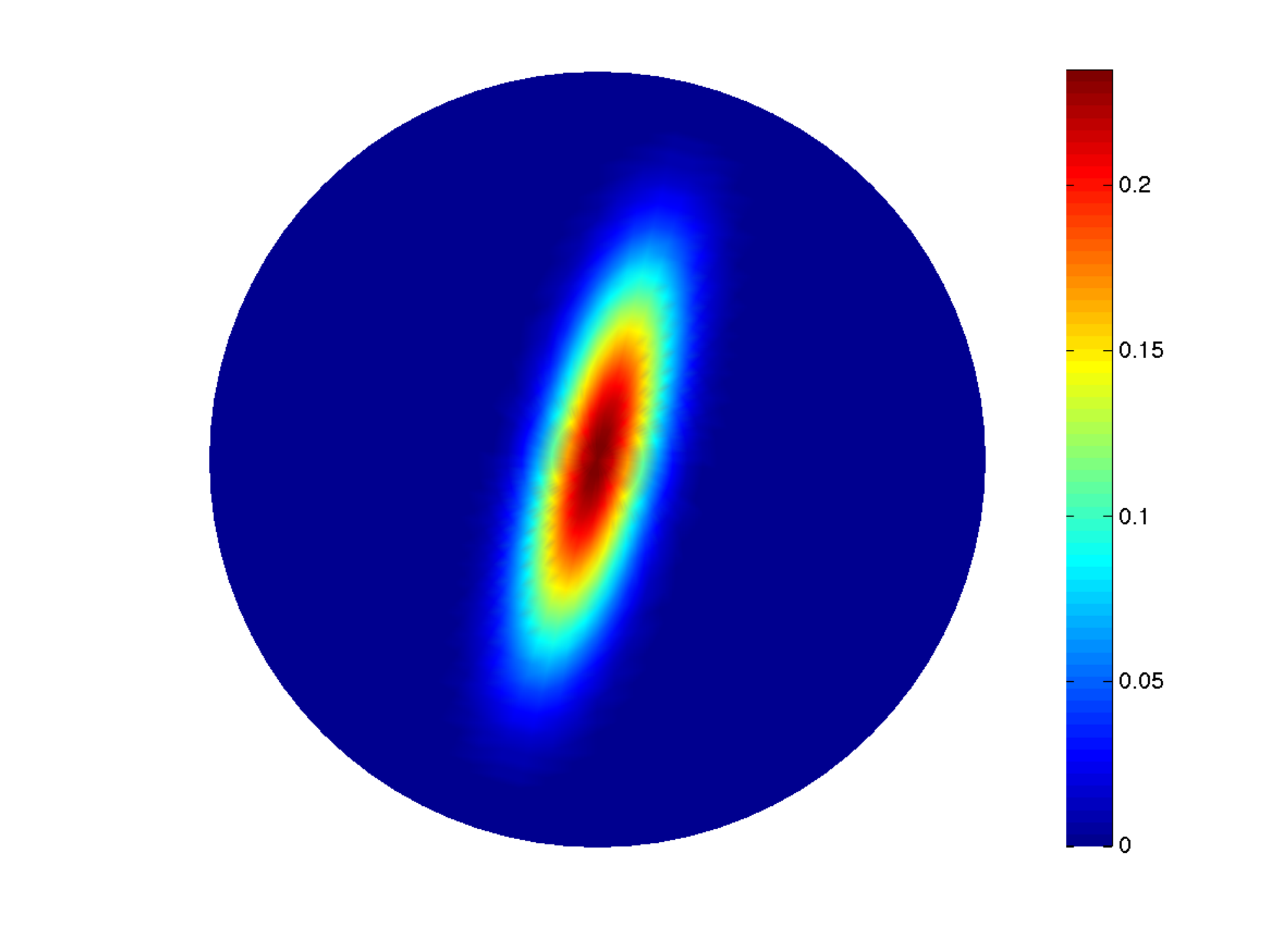}
\caption{$\Lambda t = 1.8$}
\end{subfigure}
\begin{subfigure}{0.3\textwidth}
\includegraphics[width=\textwidth]{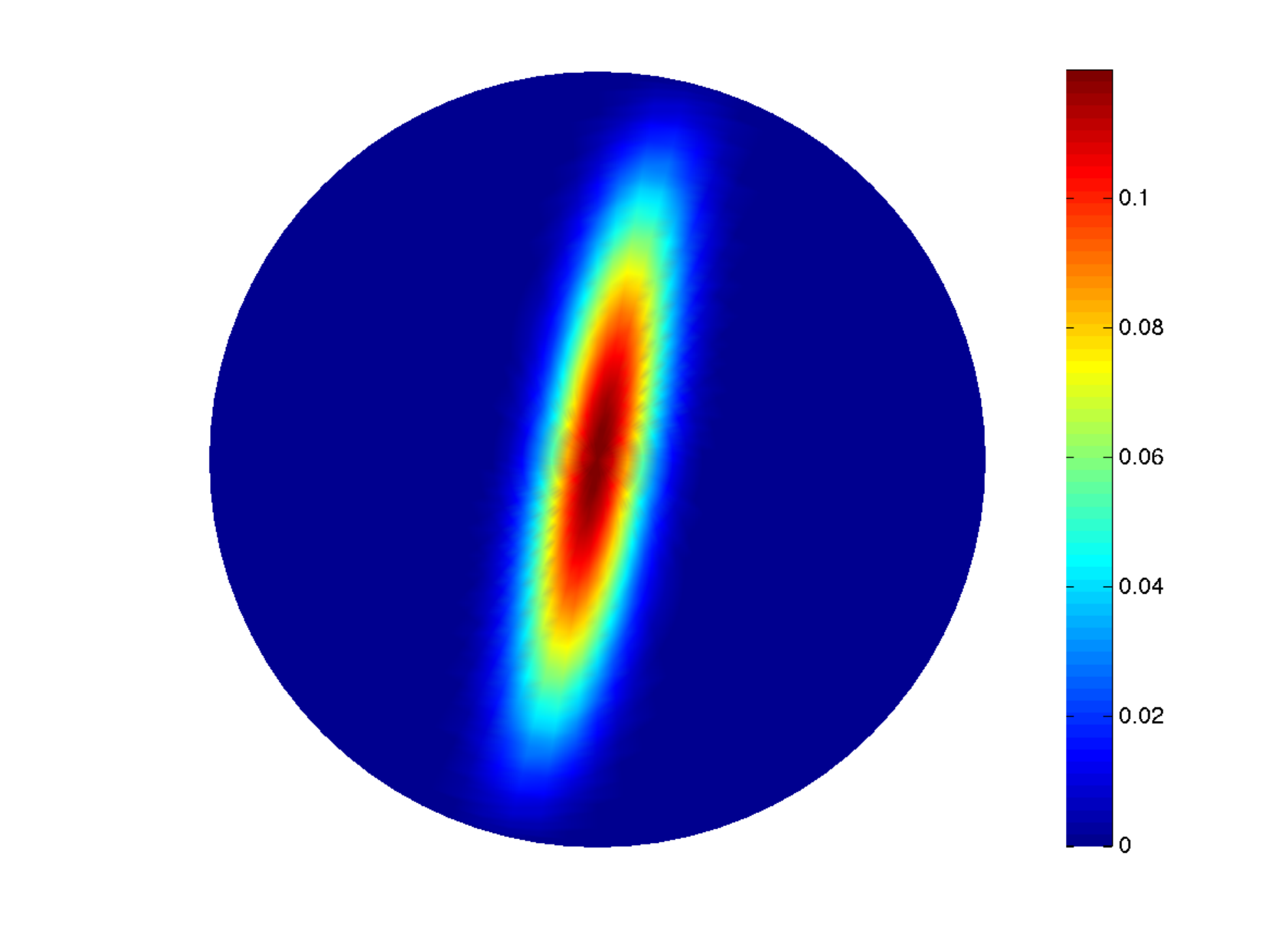}
\caption{$\Lambda t = 2.7$}
\end{subfigure}
\begin{subfigure}{0.3\textwidth}
\includegraphics[width=\textwidth]{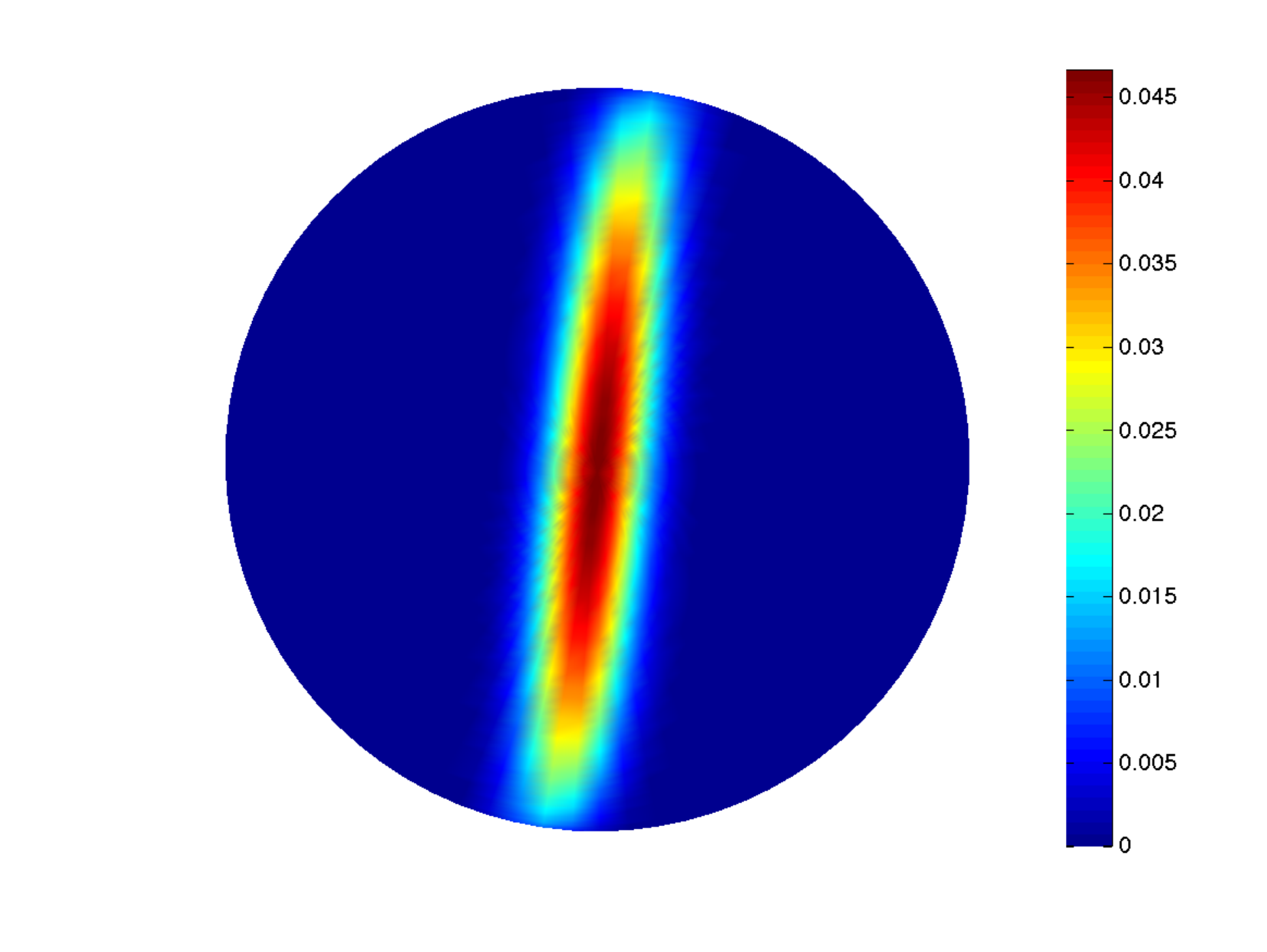}
\caption{$\Lambda t = 4.5$}
\end{subfigure}
\caption{Atomic Q-functions on the space $\set{\hat{S}_x,\hat{Q}_{yz},\hat{Q}_{zz}-\hat{Q}_{yy}}$ looking from the South Pole, for 120 atoms with $\Gamma = 0$.}

\end{figure}

\newpage

\bibliographystyle{apsrev4-1}

\end{document}